\documentclass[]{interact}

\usepackage{epstopdf}
\usepackage[caption=false]{subfig}

\usepackage[numbers,sort&compress]{natbib}
\bibpunct[, ]{[}{]}{,}{n}{,}{,}
\renewcommand\bibfont{\fontsize{10}{12}\selectfont}

\theoremstyle{plain}

\theoremstyle{definition}

\theoremstyle{remark}

\usepackage{tikz}
\usetikzlibrary{shapes.geometric}
\usetikzlibrary{shapes.arrows}
\usetikzlibrary{positioning,fit,calc}
\usepackage{multirow}

\begin{document}

\articletype{ORIGINAL ARTICLE}

\title{A feasibility study on the use of low-dimensional simulations for database generation in adaptive chemistry approaches}

\author{
\name{Ashish~S. Newale\textsuperscript{a,b}\thanks{CONTACT Ashish~S. Newale. Email: ashish.newale@gmail.com}, Pushan Sharma\textsuperscript{a}, Stephen~B. Pope\textsuperscript{a}, and Perrine Pepiot\textsuperscript{a}}
\affil{\textsuperscript{a}Sibley School of Mechanical and Aerospace Engineering, Cornell University, Ithaca, New York, USA; \textsuperscript{b}ANSYS, Inc., Lebanon, New Hampshire, USA}
}

\maketitle

\begin{abstract}
Large eddy simulation (LES)/ Probability Density Function (PDF) approaches are now well established and can be used for simulating challenging turbulent combustion configurations with strong turbulence chemistry interactions. Transported PDF methods are known to be computationally expensive compared to flamelet-like turbulent combustion models. The pre-partitioned adaptive chemistry (PPAC) methodology was developed to address this cost differential. PPAC entails an offline preprocessing stage, where a set of reduced models are generated starting from an initial database of representative compositions. At runtime, this set of reduced models are dynamically utilized during the reaction fractional step leading to computational savings. We have recently combined PPAC with in-situ adaptive tabulation (ISAT) to further reduce the computational cost. We have shown that the combined method reduced the average wall-clock time per time step of large-scale LES/particle PDF simulations of turbulent combustion by 39\%. A key assumption in PPAC is that the initial database used in the offline stage is representative of the compositions encountered at runtime. In our previous study this assumption was trivially satisfied as the initial database consisted of compositions extracted from the turbulent combustion simulation itself. Consequently, a key open question remains as to whether such databases can be generated without having access to the turbulent combustion simulation. Towards answering this question, in the current work, we explore whether the compositions for forming such a database can be extracted from computationally-efficient low-dimensional simulations such as 1D counterflow flames and partially stirred reactors. We show that a database generated using compositions extracted from a partially stirred reactor configuration leads to performance comparable to the optimal case, wherein the database is comprised of compositions extracted directly from the LES/PDF simulation itself.
\end{abstract}

\begin{keywords}
Adaptive chemistry; ISAT; LES; PDF; PaSR
\end{keywords}

\section{Introduction}

The world energy usage is forecast to grow to $50\%$ by $2050$ and a large part of this energy consumption is still projected to be met by fossil-fuel based energy sources \cite{energy2019annual}. To mitigate the adverse impact of the use of fossil-fuel based energy sources, there is a time-sensitive need for the development of innovative, high-efficiency, low-emission, combustion-based energy conversion devices. The design of such devices can be informed or derived based on predictive computations. A crucial component of predictive reacting flow simulations for cases of engineering interest is the turbulent combustion model. Turbulent combustion models can broadly be classified as flamelet-like or PDF-like \cite{pope2013small}. Transported (particle) PDF methods \cite{pope1985pdf} are the archetype of PDF-like method. These methods are deemed to be more general as do not make any assumptions on the underlying flame structure \cite{wu2015pareto}. However, this added generality comes at an increased computational cost. The increased computational cost involves significantly higher memory requirement and CPU time, even for a small hydrocarbon fuel such as methane \cite{hiremath2012computationally}.


To alleviate this cost penalty, a number of approaches have been proposed and implemented in the context of particle PDF methods. These approaches involve the use of monolithic reduced mechanisms \cite{lu2005directed,pepiot2008efficient}, dimension reduction \cite{keck1990rate,ren2006invariant}, storage retrieval \cite{pope1997computationally,franke2017tabulation}, adaptive chemistry \cite{liang2009dynamic,ren2014use,liang2015pre,d2020adaptive}, and their combinations \cite{newale2019combined,chatzopoulos2013chemistry,hiremath2011combined,contino2011coupling,ren2014use}. In this work, we focus on an adaptive chemistry and its combination with a storage retrieval approach. Specifically, we utilize the pre-partitioned adaptive chemistry (PPAC) framework developed by Liang et al. \cite{liang2015pre}. Newale et al. \cite{newale2019combined} combined PPAC with in-situ adaptive tabulation, which showed encouraging results for a partially stirred reactor (PaSR) test case. Furthermore, Newale et al. \cite{newale2020comp} developed and implemented a holistic integration of PPAC-ISAT with a LES/particle PDF solver. For a large-scale LES/PDF simulation of Sandia flame D, the average wall-clock time per time step was reduced by $39\%$ with the use of PPAC-ISAT over the use of ISAT with the detailed mechanism with minimal loss of accuracy. This showed the utility of such a combined approach to enable computationally efficient and accurate LES/PDF simulations of turbulent combustion.

PPAC entails the generation of reduced models in an offline preprocessing stage. Specifically, PPAC uses an initial database of compositions that is assumed to be representative of compositions encountered at runtime. The compositions in the database are partitioned into a user-specific number of regions such that compositions belonging to the same region share similar chemical kinetic characteristics. A reduced mechanism is then generated for each region using compositions from the database that lie within it. At runtime, an encountered composition is initially classified to identify the appropriate region. The region-specific reduced mechanism derived in the offline stage is then utilized for integrating the particle composition.\par

Based on this discussion, we note that the success of PPAC is closely tied with the database used in the preprocessing stage. The control of errors incurred at runtime due to the use of PPAC can only be assured if the initial database of compositions is representative of those encountered at runtime. Newale et al. \cite{newale2020comp} used the compositions extracted from a small domain Cartesian LES/PDF simulation using the detailed mechanism for generation of the initial database. The performance of PPAC and PPAC-ISAT using this database was quantified for the same Cartesian configuration and a full-scale cylindrical configuration. This implies that the assumption of the database compositions being representative of those encountered at runtime was satisfied exactly for the small domain LES/PDF simulation, and approximately for the large-scale simulation of Sandia flame D. \par

A key open question that remains unanswered is how does one generate an initial database of compositions representative of those encountered at runtime in turbulent combustion simulations? Obvious candidates for generation of such an initial database are existing canonical 0D/1D reactors, or low-dimensional simulations. We recently developed an ISAT-based approach \cite{newale2020towards}, which was used to examine whether compositions extracted from a series of direct numerical simulations could be emulated by compositions generated from a corresponding set of low-dimensional simulations. The comparison showed that 1D counterflow flames (flamelets) and PaSR can indeed generate compositions that are representative of those encountered in turbulent combustion simulations. \par 

A similar type of conclusion can be drawn based on recent work in a separate but similar context. Specifically, there is a growing body of work utilizing low-dimensional simulations to generate samples for training artificial neural networks (ANNs) for reaction mapping regression. The trained ANNs are then utilized to predict the reaction mapping for compositions at runtime instead of directly integrating them, leading to a significant speedup in turbulent combustion simulations. The key assumption in the usage of ANNs in such a manner, similar to PPAC, is that sample compositions used in the training stage are representative of the compositions encountered at runtime in the turbulent combustion simulation. Chatzopoulos and Rigopoulos \cite{chatzopoulos2013chemistry}, Franke et al. \cite{franke2017tabulation}, and Ding et al. \cite{ding2021machine} have shown that the use of variations of 1D counterflow laminar flames or flamelets to generate samples during the training stage leads to excellent accuracy in a-posteriori turbulent combustion simulations. Wan et al. \cite{wan2020chemistry} have shown that samples generated from a PaSR-like configuration that accounts for non-adiabatic effects leads to accurate simulations of flame-wall interaction with ANNs. \par

The objective of the current work is to directly examine and compare the performance of PPAC and PPAC-ISAT in LES/PDF computations of Sandia flame D using initial databases generated based on compositions from flamelets and from PaSR simulations. We use two configurations for the simulation of Sandia flame D, a small domain Cartesian configuration and a full-scale cylindrical domain. The configurations are identical to those used in our previous work \cite{newale2020comp}, and are chosen for the same reasons. Specifically, the smaller Cartesian domain is chosen to directly quantify the errors incurred due to the use of PPAC and PPAC-ISAT with an efficiently generated database. The cylindrical configuration is used to examine the performance of PPAC-ISAT in a large-scale particle LES/PDF simulation. The rest of the paper is structured as follows: we provide a brief description of the LES/PDF solver used in this work; a more comprehensive explanation of PPAC and PPAC-ISAT; additional details on the two configurations used; followed by results and discussion.

\section{Methods}
\subsection{LES/PDF solver}
A discretely conservative variable density low Mach solver NGA \cite{desjardins2008high} is used for performing all computations. A Lagrangian particle PDF method has been implemented in NGA to solve for the one-point one-time joint density-weighted filtered PDF of species mass fractions and enthalpy ($\tilde{f}(\bm{\psi};\bm{x},t)$, where $\bm{x}$ refers to the position, $t$ is the time, and $\bm{\psi}$ denotes the composition sample space). The exact transport equation for the filtered density weighted PDF equation \cite{pope2010self} for a variable density reacting flow after neglecting the effects of participating media radiation and invoking the unity Lewis number assumption is detailed below: 
\begin{equation}
    \dfrac{\partial \bar{\rho} \tilde{f}}{\partial t}+ \nabla.[\bar{\rho}\tilde{f}(\tilde{\bm{u}}+\widetilde{\bm{u}'|\bm{\psi}})] = -\dfrac{\partial}{\partial \bm{\psi}_\alpha}\bigg[\bar{\rho} \tilde{f}\big(\dfrac{1}{\bar{\rho}} \overline{\nabla.(\rho\Gamma\nabla\phi_{\alpha}|\bm{\psi})} \bigg] -\dfrac{\partial (\bar{\rho} \tilde{f} S_\alpha(\bm{\psi}))}{\partial \psi_\alpha}. 
\end{equation}
Here, $\bm{\phi}$ refers to the composition which includes all the species mass fractions and enthalpy. $\tilde{\bm{u}}$ is the resolved velocity and $\bm{u}'$ is the residual velocity fluctuation, $\Gamma$ is the molecular diffusivity, and $S_\alpha$ is the reaction source term for species $\alpha$. We note that as the unity Lewis number assumption is used, the molecular diffusivity is identical for all species and equal to the thermal diffusivity. \par

The conditional diffusion term (first term on the right hand side) is unclosed as it involves a gradient of the species mass fraction. This term is closed using the interaction by exchange with mean (IEM) model \cite{villermaux1972representation} as follows: 
\begin{equation}
    \dfrac{1}{\bar{\rho}} \overline{\nabla.(\rho\Gamma\nabla\phi_{\alpha}|\bm{\psi})} = -\Omega_m(\psi_{\alpha}-\tilde{\phi}_{\alpha}),
\end{equation}
where $\Omega_m$ is determined as follows: 
\begin{equation}
      \Omega_m = C_m \dfrac{\tilde{\Gamma}+\tilde{\Gamma_t}}{\Delta^2},
\end{equation}
where $\Gamma_t$ is the turbulent diffusivity, $C_m$ is the mixing model constant, and $\Delta$ is LES filter width. 

Additionally, the residual velocity fluctuation conditioned on the scalar is closed using the gradient diffusion approximation. The remaining terms including the chemical reaction source term appear in closed form. \par

The modeled PDF equation is solved using a particle Monte Carlo scheme \cite{pope1985pdf}, where notional particles evolve in the physical and composition space. The notional particles evolve in physical space according to the following stochastic differential equation (SDE):
\begin{equation}
    d\bm{X}^*(t) = \bigg(\tilde{\bm{u}} + \dfrac{\nabla \bar{\rho}(\tilde{\Gamma}+\tilde{\Gamma_t})}{\bar{\rho}}\bigg)^*dt+\sqrt{2(\tilde{\Gamma}^*+\tilde{\Gamma_t}^*)}d\bm{W},
    \label{eq:1a}
    \end{equation}
    where $\bm{X}^*$ is the particle position and $\bm{W}(t)$ is an isotropic Wiener process. The superscript `$*$' refer to quantities evaluated at the particle location.  
    
The particles evolve in composition space due to mixing and reaction as follows:
    \begin{equation}
    d\bm{\phi}^*(t) = -\Omega_m^*(\bm{\phi}^*-\tilde{\bm{\phi}}^*)dt + S(\bm{\phi}^*)dt.
    \label{eq:2}
\end{equation}

The particle SDEs are integrated using a simple first-order splitting scheme. This entails completing in order transport, mixing, and reaction fractional steps. As the names suggest, the transport fractional step updates the particle position in physical space according to Eq. \ref{eq:1a}, the mixing fractional step handles changes occurring in particle compositions due to molecular mixing (first term of Eq. \ref{eq:2}), and the reaction fractional step advances particle compositions to account for chemical reactions (second term of Eq. \ref{eq:2}). The transport fractional step is performed using a simple forward Euler method. As noted before, the mixing fractional step uses the IEM model \cite{villermaux1972representation}. The reaction fractional step is performed efficiently using a dynamic load balancing strategy \cite{gruselle2014etude}. 

The mean estimation and interpolation to particle location is performed using a cloud-in-cell linear spline method \cite{viswanathan2011numerical}. Additionally, a smoothing strategy is utilized for variance reduction \cite{viswanathan2011numerical}. The LES/PDF solver developed is two-way coupled. This implies that the LES solver uses a resolved density computed based on the PDF solution. Additionally, the resolved composition from the PDF solver is used to compute the kinematic viscosity and molecular diffusivity. The resolved velocity and turbulent diffusivity from the LES solver is used to advance the particle positions. The direct use of the density computed using the PDF solver for advancing LES fields can lead to numerical instabilities. Consequently, we use the transported specific volume (TSV) approach of Popov et al. \cite{popov2015specific} for handling the density coupling between LES and PDF solvers. 



\subsection{Pre-partitioned adaptive chemistry (PPAC)}
\tikzstyle{startstop} = [rectangle, rounded corners, minimum width=3cm, minimum height=1cm,text centered, draw=black]
\tikzstyle{tmp} = [diamond, rounded corners, minimum width=3cm, minimum height=1cm,text centered, draw=black]
\tikzstyle{arrow} = [thick,->,>=stealth]
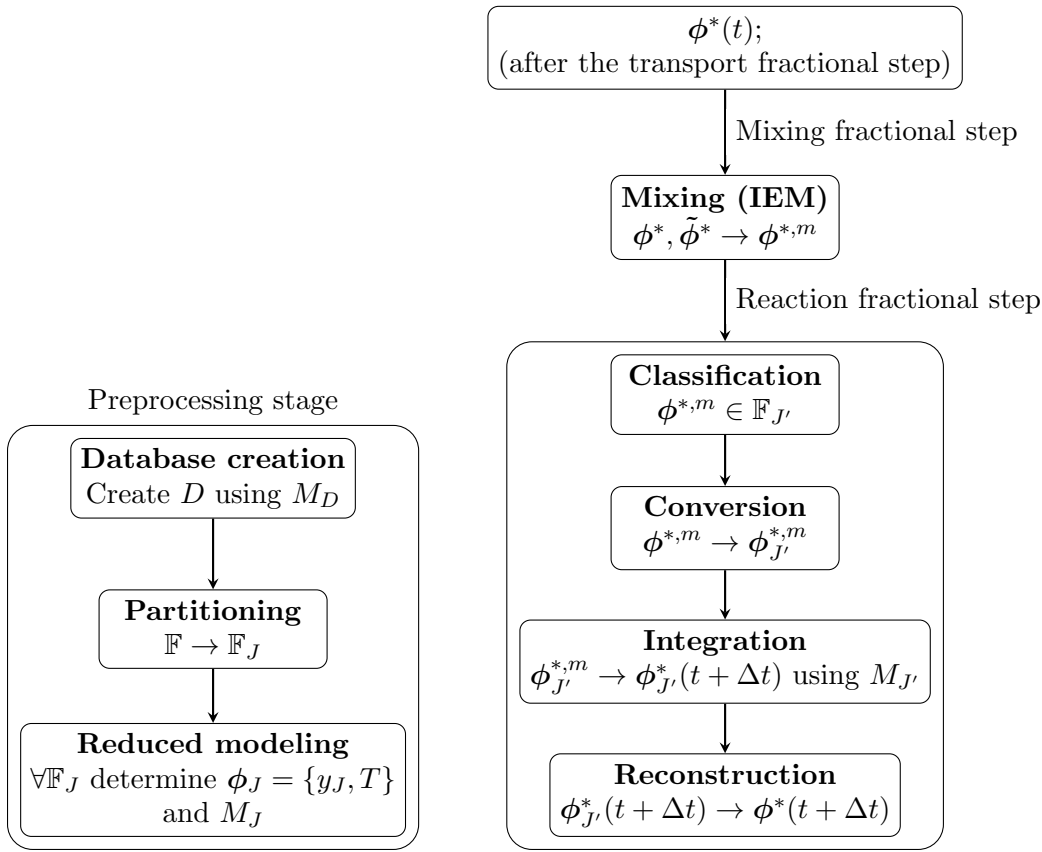
\begin{figure}  
	\centering
	\begin{tikzpicture}[node distance=2cm]
	\node (start) [startstop][draw, align=center] {\textbf{Database creation}\\
		Create $D$ using $M_D$};
	\node (partitioning) [startstop, below of=start][draw, align=center] {\textbf{Partitioning}\\
		$\mathbb{F} \rightarrow \mathbb{F}_J$};	    	 
	\node (rm) [startstop, below of=partitioning][draw, align=center] {\textbf{Reduced modeling}\\
		$\forall \mathbb{F}_J$ determine $\bm{\phi}_J = \{y_J,T\}$\\ and $M_J$};   
	\node [draw=black!120!black,rectangle, rounded corners=2ex,
	align=center, inner sep=1ex, label={Preprocessing stage}, fit={(start) (partitioning) (rm)}] {};
	\draw [arrow] (start) -- (partitioning);                  
	\draw [arrow] (partitioning) -- (rm);   
	\end{tikzpicture}
	\qquad
	\begin{tikzpicture}[node distance=2.25cm,auto]
	\node (start) [startstop][draw, align=center] {$\bm{\phi}^{*}(t)$; \\
		(after the transport fractional step)};
	

	\node (mix) [startstop, below of=start][draw, align=center] {\textbf{Mixing (IEM)}\\
		$\bm{\phi}^{*}, \bm{\tilde{\phi}}^* \rightarrow\bm{\phi}^{*,m}$ };     
		
	\begin{scope}[node distance=1.77cm]
	\node (classify) [startstop, below=1.25cm of mix][draw, align=center] {\textbf{Classification}\\
		$\bm{\phi}^{*,m} \in \mathbb{F}_{J'}$};
	\node (red) [startstop, below of=classify][draw, align=center] {\textbf{Conversion}\\
		$\bm{\phi}^{*,m} \rightarrow \bm{\phi}^{*,m}_{J'}$}; 
	\node (di) [startstop, below of=red][draw, align=center] {\textbf{Integration}\\
		$\bm{\phi}^{*,m}_{J'} \rightarrow \bm{\phi}^{*}_{J'}(t+\Delta t)$ using $M_{J'}$};
		
	\node (recon) [startstop, below of=di][draw, align=center] {\textbf{Reconstruction}\\
	$\bm{\bm{\phi}}^{*}_{J'}(t+\Delta t)\rightarrow\bm{\phi}^{*}(t+\Delta t)$};
	 
	\end{scope}
	\node (reacf)[draw=black!120!black,rectangle, rounded corners=2ex,
	align=center, inner sep=1ex, fit={(classify) (red) (di) (recon)}] {};
	\draw [arrow] (start) -- (mix)node[midway,fill=white]{Mixing fractional step};                  
	\draw [arrow] (mix) -- (reacf)node[midway,fill=white]{Reaction fractional step}; 
	\draw [arrow] (classify) -- (red);
	\draw [arrow] (red) -- (di);
	\draw [arrow] (di) -- (recon);
	\end{tikzpicture}
	\caption{The offline (left) and online (right) stages of the PPAC methodology.}
	\label{fig:sc}
\end{figure}   

As noted previously, PPAC \cite{liang2015pre} involves an offline preprocessing stage and a runtime stage. This section provides key details for both stages. A more comprehensive description of the steps involved in each of these stages has been provided by Liang et al. \cite{liang2015pre}. The offline preprocessing stage utilizes a database of compositions to partition the composition space into a user-specified number of regions. Region-specific reduced mechanisms and corresponding reduced representations are derived subsequently for each region. A schematic of the offline stage is shown in the left half of Fig. \ref{fig:sc}. $D$ is the initial database of compositions, $M_D$ refers to the detailed chemical kinetic mechanism. $J$ and $J'$ refer to two of the user-specified number of regions created by partitioning the composition space. $\phi_J$ is the reduced skeletal representation and $M_J$ is the reduced model for region $J$. The superscript $*$ implies quantities computed at the particle location, and $m$ is used to show the composition after the mixing fractional step. The simulation time step is denoted by $\Delta t$. The offline stage involves the following steps:
\begin{itemize}
    \item \textbf{Database creation}: A set of detailed compositions ($D$) that are representative of compositions encountered at runtime is assembled. The compositions from the database are assumed to densely sample the composition space accessed by the turbulent combustion simulation. It is important to appreciate that this is a crucial assumption as it implies that the accuracy of PPAC at runtime is directly linked to the initial database being representative of compositions encountered at runtime. 
    \item \textbf{Partitioning}: The compositions from the database are partitioned into a user-specified number of regions such that the compositions belonging to the same region share similar chemical kinetic characteristics. The partitioning is performed using a recursive hyperplane cutting algorithm, using a DRGEP coefficient based metric to quantify the homogeneity of kinetic characteristics \cite{liang2015pre}. Additionally, we note that the partitioning is not done directly in the full composition space ($\mathbb{F}$) but in a classifying space with a significantly smaller number of species \cite{liang2015pre}. The species used to define the classifying space ($\mathbb{C}$) are identified using principal component analysis (PCA) \cite{yang2013empirical}. 
    \item \textbf{Reduced modeling}: A reduced model ($M_J$) is generated for each region using the directed relation graph with error propagation (DRGEP) \cite{pepiot2008efficient} using compositions from the database that lie within the region. Finally, given a user-specified reduction threshold, a reduced model with the smallest number of species is selected for each region such that the resulting reduction reaction mapping errors are less than or equal to the reduction threshold. We note here that the selection of reduced models is done on the basis on the reduction reaction mapping errors as opposed to the use of DRGEP threshold in dynamic adaptive chemistry methods \cite{liang2009dynamic}. This is crucial as the use of a specific DRGEP threshold does not imply that the reduction reaction mapping errors will be bounded by known values that can be easily computed \cite{xie2018dynamic}. Hence, the direct use of reduction reaction mapping errors for choosing reduced mechanisms in the offline stage leads to bounded reduction reaction mapping errors in the online stage, if the database compositions are representative of those encountered at runtime. 
\end{itemize}

A schematic of the online stage of PPAC is provided in Fig. \ref{fig:sc}. In the current implementation, PPAC specific operations are performed only in the reaction fractional step. Consequently, the current implementation is not limited to the IEM model used here, and is compatible with other mixing models. The details of the reaction fractional step are provided below: 

\begin{itemize}
    \item \textbf{Classification}: At the start of the reaction fractional step, the region to which the particle composition belongs ($J'$) is identified using an efficient low-dimensional binary-tree search . 
    \item \textbf{Reduction}: The particle composition is converted to its region-specific skeletal representation ($\bm{\phi}^{*,m}_{J'}$). 
    \item \textbf{Integration}: The particle composition in its skeletal representation is integrated using the region-specific reduced mechanism derived in the offline stage. 
    \item \textbf{Reconstruction}: The integrated particle composition is converted to its detailed representation ($\bm{\phi}^{*}(t+\Delta t)$). 
\end{itemize}

Note that the detailed representation does not refer to the use of a state vector corresponding to the full detailed mechanism. Further details on what we refer to as the detailed representation are provided in section \ref{sec:num_impl}.

\subsection{PPAC-ISAT}
The formulation for PPAC-ISAT is identical to that of PPAC with one key distinction. As opposed to directly integrating the particle compositions, the particle compositions are resolved using ISAT \cite{lu2009improved}. As the particle compositions are resolved in their skeletal representation, a region-specific ISAT table is maintained on each core. The resolution of compositions in their skeletal representation leads to faster build and retrieve times as the tabulation is performed in a reduced number of dimensions. Additionally, the use of reduced mechanisms leads to more efficient integration for operations such as adds, grows, and direct evaluations \cite{newale2019combined}. 

\subsection{Numerical implementation}
\label{sec:num_impl}
We use the same implementation detailed in our previous work \cite{newale2020comp}. The locally significant species \cite{liang2015pre} are neglected. We have shown in past work that this does not alter the results appreciably \cite{newale2020comp}. Hence, the particles only retain the species that are included in at least one reduced model. The species that are not included in any region-specific reduced model are eliminated from the particle species mass fraction vector. Additionally, operations such as mean estimation are performed very efficiently as the indices of species with non-zero mass fractions are known \textit{a-priori}.

\section{Simulation details}



\subsection{LES/PDF configurations}
We utilize a Cartesian configuration which extends $18D$ in the streamwise direction and $10D$ in the cross-stream directions, where $D$ is the diameter of the fuel nozzle. This configuration uses a uniform $90$ (streamwise) $\times 100 \times 100$ mesh. This configuration is similar to that of Sheikhi et al. \cite{sheikhi2005large} and identical to the Cartesian configuration used in our recent work \cite{newale2020comp}. As in our previous work \cite{newale2020comp}, $C_m$ is specified to be $8$ based on an examination of the resolved statistics at $x/D=15$. The modest size of this Cartesian domain enables the direct quantification of errors incurred due to the use of PPAC and PPAC-ISAT. 

A full-scale cylindrical configuration extends to $60D$ in the axial direction, $20D$ in the radial direction, and $2\pi$ in the azimuthal direction. A non-uniform $192$ (axial) $\times 128$ (radial) $\times 32$ (azimuthal) mesh is used for this configuration. The mixing model constant ($C_m$) is specified to be 4. The specification of the mixing model constant is based on a sensitivity study of the resolved mean and RMS of species mass fraction and temperature \cite{newale2020comp}. Both the Cartesian and cylindrical configurations nominally utilize $25$ particles per cell. 

\subsection{PPAC input specifications}
The parameters utilized for PPAC are identical to those used in our recent work \cite{newale2020comp}, to enable a direct comparison of the results. As before, a detailed $38$ species mechanism \cite{esposito2011skeletal} is used in this work. Liang et al. \cite{liang2015pre} have shown that the choice of the number of regions, provided that it is large enough, does not significantly impact the performance of PPAC. Here, the number of regions is set to 10, a number found large enough in our previous work \cite{newale2020comp}. This number is also small enough to allow for detailed analysis of the results. The targets for DRGEP reduction are specified as CH$_4$, CO, OH, HO$_2$, and heat release rate. This set of targets is the same as our previous work \cite{newale2020comp}, and has been shown to be effective in capturing the relevant kinetics. The use of an identical set of targets also enables a consistent comparison with the results for a close-to-optimal database case we had shown previously \cite{newale2020comp}. The reduction thresholds examined for the Cartesian case include $5\times 10^{-4}$, $10^{-4}$, $5\times10^{-5}$, $10^{-5}$, and $10^{-6}$. Only the reduction threshold of $10^{-5}$ is explored for the full-scale cylindrical configuration. \par

We explore two different low-dimensional simulations for generation of the initial database: 1D counterflow non-premixed flames (flamelets) and PaSR \cite{pope1997computationally}. The stream compositions for the flamelets match the fuel and coflow compositions for flame D. The flamelets are computed using the FlameMaster code \cite{pitsch1994flamemaster} in mixture fraction space for increasing scalar dissipation rates till extinction. The PaSR uses three streams, with the fuel, oxidizer, and pilot streams matching the corresponding compositions for flame D. The relative stream mass flow rate for the pilot is set to be 0.1. The fuel and oxidizer relative stream mass flow rates are set to make the overall reactor stoichiometric. The particles are mixed using a pairwise mixing model. Both the pairing and mixing time are $1$ ms and the time step used is $0.1$ ms. A comprehensive description of the PaSR implementation used here is provided in Liang et al. \cite{liang2015pre}.

\subsection{ISAT and PPAC-ISAT table parameter specifications}
The maximum table size for the use of ISAT with the detailed mechanism is set to $500$ MB. The region-specific ISAT tables are restricted to a maximum size of $50$ MB. The ISAT error tolerance is specified as $10^{-4}$. This ISAT error tolerance is chosen as it leads to acceptable tabulation errors for the standalone ISAT case \cite{hiremath2012computationally} and the combined PPAC-ISAT case \cite{newale2020comp}.

\section{Results and discussion}
As discussed in the previous section, we utilize two different configurations. The smaller Cartesian configuration is used to directly quantify the errors incurred due to the use of PPAC and PPAC-ISAT, while the cylindrical configuration is used to test the performance of PPAC-ISAT in a large-scale LES/PDF simulation. We initially present the results for the Cartesian configuration, followed by details on the performance of PPAC-ISAT for the cylindrical configuration.

\subsection{Cartesian configuration}
\label{ssec:cartesian}
The operator used for converting a particle composition from its full representation to its skeletal representation does not conserve elemental mass fractions \cite{liang2015pre} as information is lost in the conversion. Hence, it is important to check the conservation errors incurred due to the use of PPAC and PPAC-ISAT. The normalized conservation error ($\hat{\varepsilon}_X$) for a quantity $X$ is computed as follows:
\begin{equation}
 \hat{\varepsilon}_X = \dfrac{1}{n_t} \sum\limits_{k=1}^{n_t} \dfrac{\sum_{n=1}^{n_p}(X_{k+}^{(n)}-X_k^{(n)})}{\sum_{n=1}^{n_p}X^{(n)}_k},
\end{equation}
where $X_k^{(n)}$ and $X_{k+}^{(n)}$ denote the value of the quantity $X$ at time step $k$ for particle $n$ before and after the mixing and reaction fractional step respectively, $n_p$ is the total number of particles, and $n_t$ is the total number of time steps. The normalized elemental conservation errors are computed from LES/PDF simulations run for two flow-through times determined based on the jet inlet velocity, starting from a statistically stationary state. We examine five different reduction thresholds for this configuration, which correspond to five different PPAC/PPAC-ISAT LES/PDF simulation runs. \par

\begin{figure}
 	\centering
 	\subfloat[Flamelet database]{%
 		\resizebox*{0.4975\textwidth}{!}{\begin{tikzpicture}
		\node[inner sep=0pt] (A) {\includegraphics{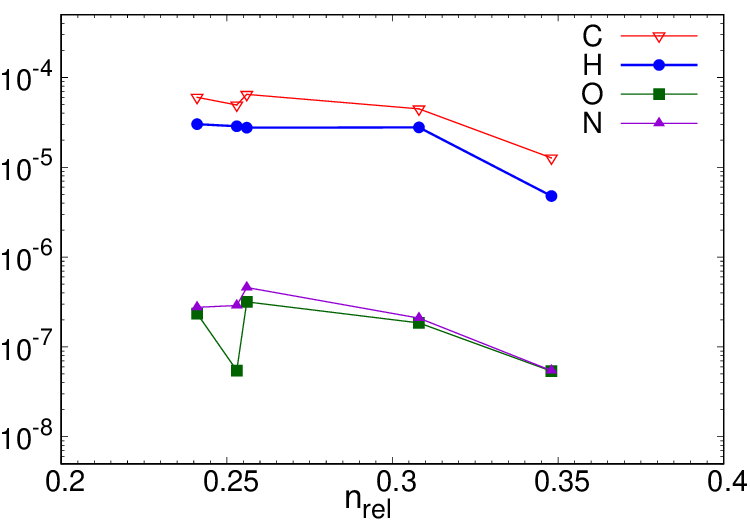}};
		\node[black,rotate=90,font=\fontsize{21}{0}\selectfont] (C) at ($(A.west)!-.035!(A.east)$) {$\mathbf{\hat{\varepsilon}_{X}}$};
		\end{tikzpicture}}
 		\label{cons_flamelet}}
 	\subfloat[PaSR database]{%
 		\resizebox*{0.4975\textwidth}{!}{\begin{tikzpicture}
		\node[inner sep=0pt] (A) {\includegraphics{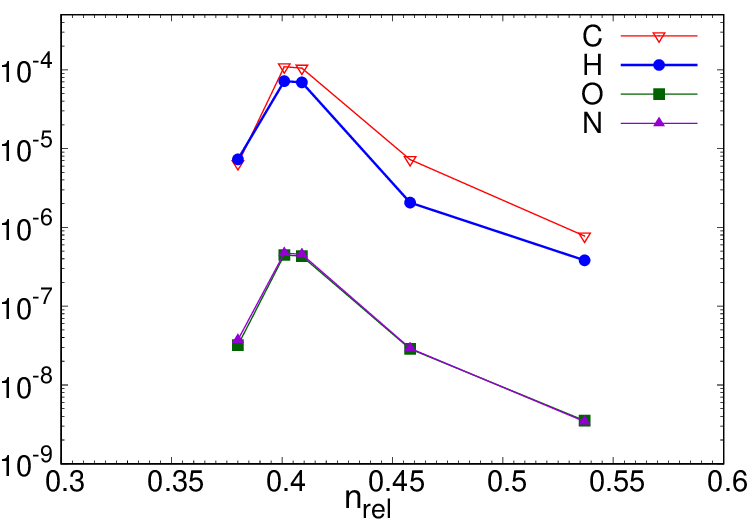}};
		\node[black,rotate=90,font=\fontsize{21}{0}\selectfont] (C) at ($(A.west)!-.035!(A.east)$) {$\mathbf{\hat{\varepsilon}_{X}}$};
		\end{tikzpicture}}
 		\label{cons_pasr}}
 	\caption{The elemental conservation errors for PPAC-ISAT with two different databases as a function of the relative number of species. Note that the range of the axes in the two plots is not identical.}
 	\label{fig_cons}
 \end{figure}

The elemental conservation errors as a function of the relative number of species for PPAC-ISAT are shown in Fig. \ref{fig_cons}. For brevity, similar plots for PPAC are not shown here. The relative number of species quantifies the average number of species included in the region-specific reduced mechanism used for integrating particle compositions, and is computed as follows \cite{liang2015pre}: 
\begin{equation}
    n_{rel} = \dfrac{1}{n_t n_p n_s} \sum\limits_{k=1}^{n_t} \sum\limits_{n=1}^{n_p} n_{s,k}^{(n),A},
\end{equation}
where $n_{s,k}^{(n),A}$ is the number of species in the reduced model used for integrating the composition of particle $(n)$ at time step $k$, and $n_s$ is the number of species in the detailed mechanism. Two distinct plots are shown corresponding to the two different low-dimensional simulations, flamelets and PaSR, for the generation of the initial database used in the offline preprocessing stage. We observe that the general trend is similar in both cases. Specifically, we observe that decreasing normalized elemental conservation errors correlate with an increase in the relative number of species. We note that an increase in the relative number of species implies the use of larger reduced mechanisms for integration of particle compositions. The use of larger reduced mechanisms leads to a reduction in the loss of information in the conversion from the full to the skeletal representation and consequently, reduced conservation errors. The conservation errors for both sets of the simulations are approximately bounded by $10^{-4}$. These errors are deemed to be acceptable. Comparing the results for the two sets of simulations, we observe that the relative number of species for the set of simulations using PPAC models generated using the PaSR database are higher than that for the models generated using the flamelet database. This difference is largely attributed to the difference in the size of the reduced model used for resolving unmixed coflow compositions, which constitute a large fraction of the total number of particles. More specifically, the reduced model obtained using the PaSR database contains a larger number of species compared to the corresponding model obtained using the flamelet database. 


\begin{figure}[t]
 	\centering
 	\subfloat[Flamelet database]{%
 		\resizebox*{0.4975\textwidth}{!}{\includegraphics{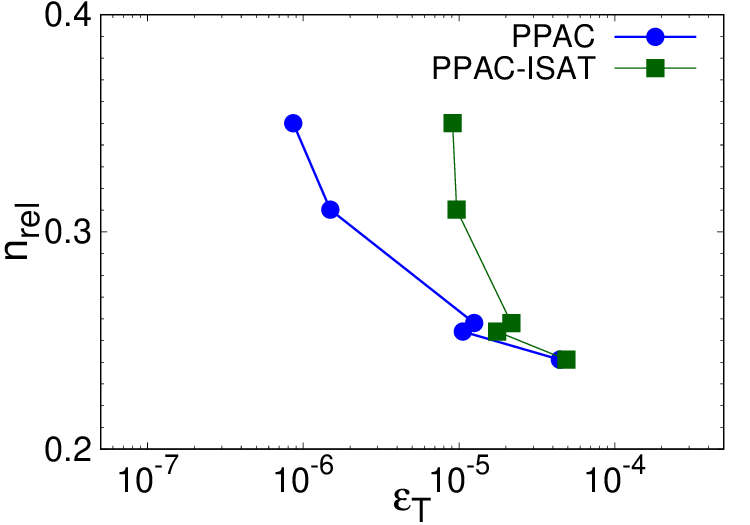}}
 		\label{inc_flamelet}}
 	\subfloat[PaSR database]{%
 		\resizebox*{0.4975\textwidth}{!}{\includegraphics{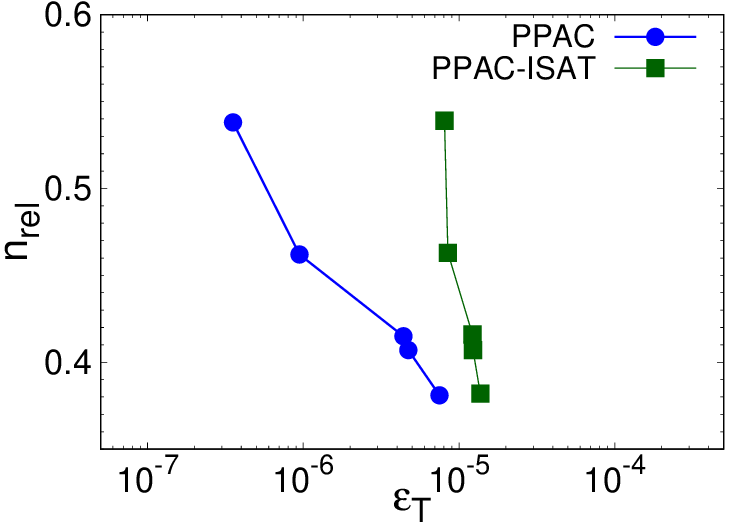}}
 		\label{inc_pasr}}
 	\caption{The relative number of species as a function of the incurred error in temperature for PPAC and PPAC-ISAT with two different databases. Note that the range of the axes in the two plots is not identical.}
 	\label{fig_inc}
 \end{figure}

Next, we examine the errors incurred due to the use of PPAC and PPAC-ISAT using the flamelet and PaSR databases. Figure \ref{fig_inc} shows the relative number of species as function of the incurred error in temperature. The incurred errors for species mass fractions with PPAC are known to be proportional to the incurred error in temperature \cite{liang2015pre}. Hence, the incurred error in temperature is deemed to be adequate for characterization of incurred errors.  The incurred error ($\varepsilon_X$) for a quantity $X$ is computed as follows: 
\begin{equation}
    \varepsilon_X = \dfrac{\sum_{k=1}^{n_t}\sum_{n=1}^{n_p}\big|X^{(n),A}_k - X^{(n),D}_k\big|}{\sum_{k=1}^{n_t}\sum_{n=1}^{n_p}|X^{(n),D}_k|},
\end{equation}
where, $X^{(n),A}_k$ and $X^{(n),D}_k$ are the integrated values for quantity $X$ of particle $(n)$ at time step $k$ obtained using PPAC/PPAC-ISAT and with direct integration using the detailed mechanism respectively. We note that the use of this metric necessitates that each particle composition in the LES/PDF simulation is integrated twice at each time step, once using PPAC/PPAC-ISAT and then using the detailed mechanism. The incurred errors are obtained from LES/PDF simulations run for $125$ time steps, starting from the statistically stationary state. The use of this metric is enabled by the relatively modest size of the computational domain for this configuration.

For PPAC in both flamelet and PaSR cases, we observe that an increase in the relative number of species leads to a decrease in the incurred error in temperature.  This is expected as the increase in the relative number of species is caused by the use of progressively less reduced mechanisms, which lead to a reduction in the incurred error in temperature. PPAC-ISAT incurs both reduction and tabulation errors due to the use of region-specific reduced mechanisms via PPAC and ISAT respectively. PPAC on the other hand only incurs reduction errors. We expect that the tabulation errors incurred due to the use of ISAT are approximately equal for the PPAC-ISAT simulations using different reduction thresholds. Additionally, the reduction errors at the same reduction threshold for PPAC and PPAC-ISAT are expected to be identical. Hence, the relative contributions of the reduction and tabulation errors in PPAC-ISAT can be determined by comparing the incurred error in temperature for PPAC and PPAC-ISAT at the same reduction threshold, which leads to the same relative number of species. Specifically, if the reduction error dominates, PPAC and PPAC-ISAT should lead to approximately similar incurred errors in temperature. If the tabulation error dominates the reduction error, we should observe distinctly different values for the incurred error in PPAC and PPAC-ISAT. Based on these arguments, we can infer for the simulations using the flamelet database that the reduction error dominates for the three largest reduction thresholds, while the tabulation error dominates for the two more stringent reduction thresholds. For the simulation using the PaSR database, we observe that the tabulation error dominates for all the reduction thresholds. This is due to the lower reduction errors for the simulations using the PaSR database compared to those using the flamelet database, especially at the larger reduction thresholds. 


\subsection{Cylindrical configuration}

The incurred error in temperature for PPAC-ISAT in the Cartesian configuration at a reduction threshold of $10^{-5}$, for both database cases is approximately $10^{-5}$. This magnitude of the incurred error in temperature has been observed to lead to minimal loss of accuracy in the resolved statistics for the full-scale cylindrical configuration \cite{newale2020comp}. Hence, the results for the full-scale cylindrical configuration are obtained with a reduction threshold of $10^{-5}$ with PPAC-ISAT. Additionally, this reduction threshold is identical to that used in our previous work \cite{newale2020comp} for the same cylindrical configuration, thereby enabling a justifiable comparison.

\begin{figure}
 	\centering
 		\resizebox*{0.4875\textwidth}{!}{\includegraphics{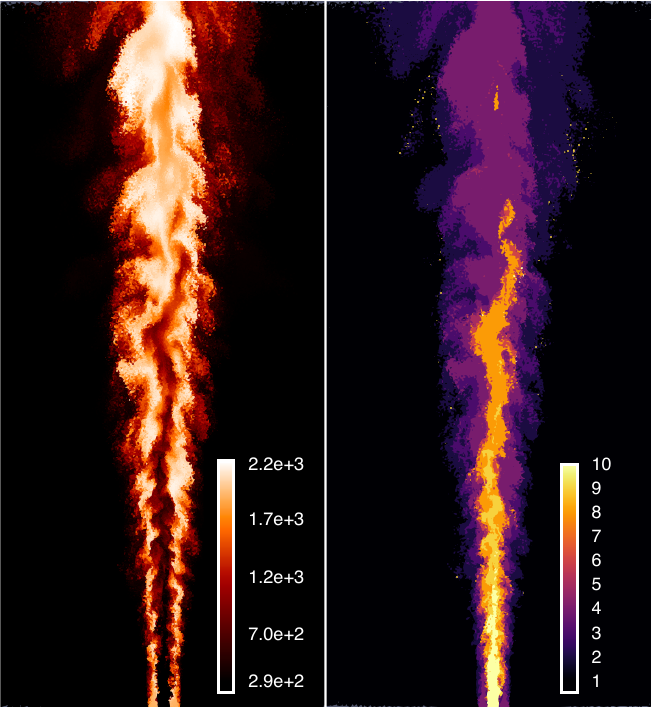}}
 		\label{contour_flamelet}
 		\resizebox*{0.4875\textwidth}{!}{\includegraphics{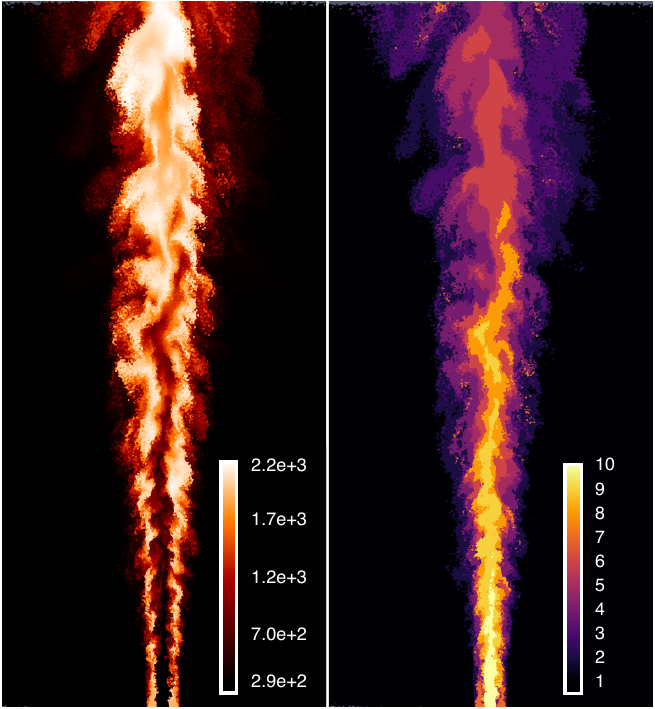}}
 		\label{contour_pasr}\\
 		\subfloat[Flamelet database]{%
 		\resizebox*{0.4975\textwidth}{!}{\includegraphics{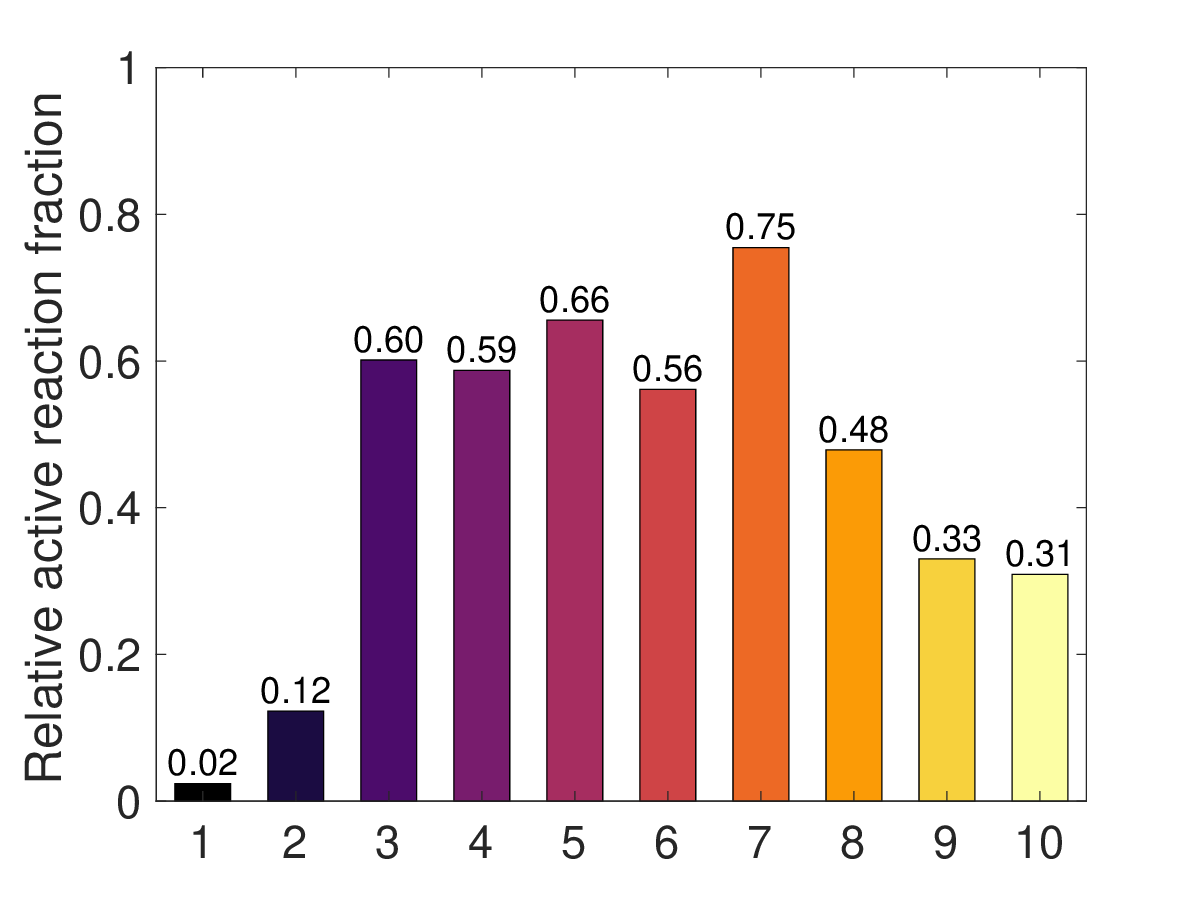}}
 		\label{bar_flamelet}}
 	\subfloat[PaSR database]{%
 		\resizebox*{0.4975\textwidth}{!}{\includegraphics{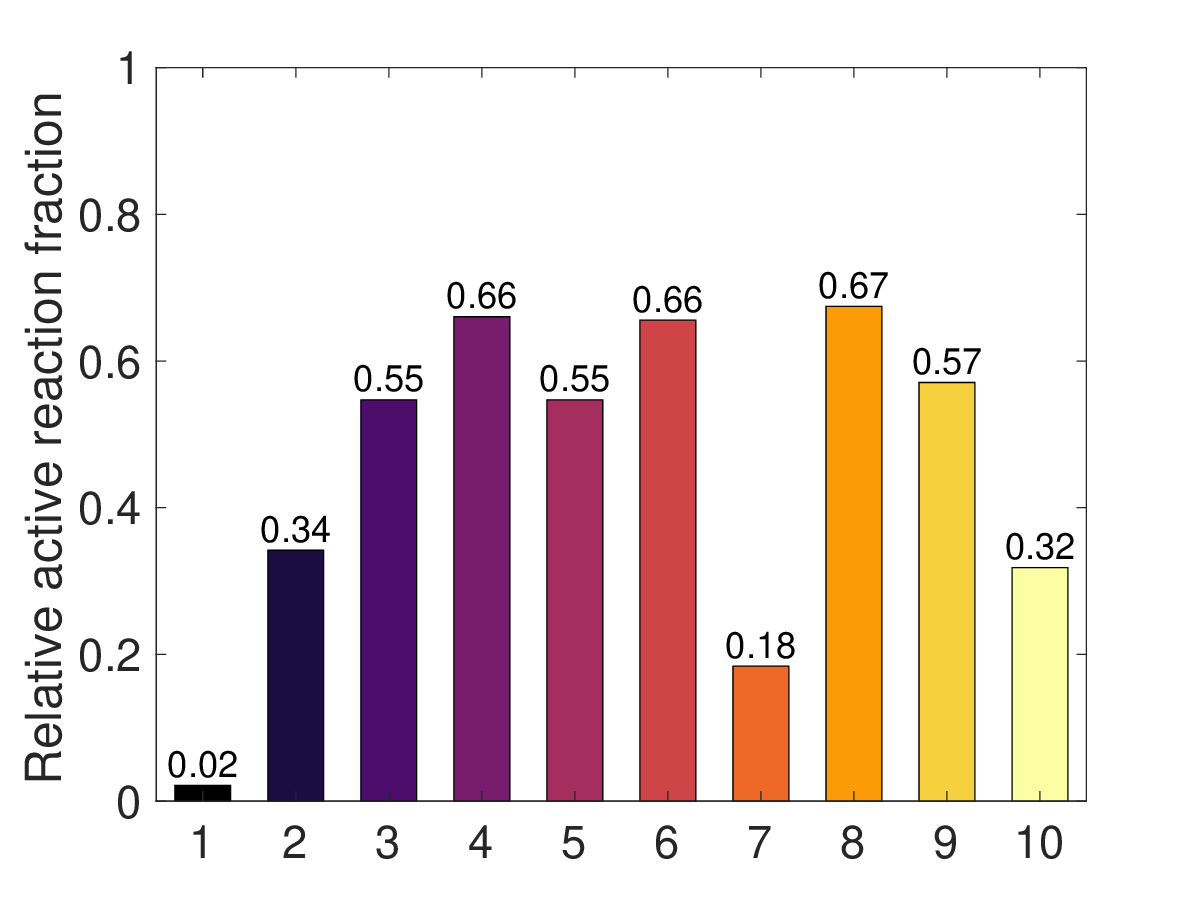}}
 		\label{bar_pasr}}
 	\caption{A cross-sectional view of the instantaneous PDF particle distribution colored by temperature (K) and model ID (top), and a histogram for the relative active reaction fraction for each region-specific reduced model (bottom). The results for the full-scale PPAC-ISAT LES/PDF simulation using reduced models generated with the flamelet and PaSR databases are shown on the left and right respectively.}
 	\label{qualtitative_comparison}
 \end{figure}

Figure \ref{qualtitative_comparison} shows the instantaneous PDF particle temperature distributions and the corresponding model IDs for both the full-scale LES/PDF PPAC-ISAT simulation using reduced models generated with the flamelet (left) and PaSR (right) databases. To provide further details on the size of these reduced models, the bottom half of the figure shows histograms for the relative active reaction fraction corresponding to these model IDs for the flamelet and PaSR databases on the left and right respectively. The relative active reaction fraction is simply the number of reactions included in the region-specific reduced model normalized by the number of reactions in the detailed mechanism. The colors for the model IDs shown in the cross-sectional view match the colors used in the histogram to facilitate correlating size of the model and its usage at various spatial locations within the domain. We observe that for both the flamelet and PaSR databases, the compositions in the unmixed coflow are handled by a model with a very small number of reactions. The compositions in the potential core of the fuel jet are also resolved with the region-specific reduced models containing only $31\%$ and $32\%$ of the reactions in the detailed mechanism for the flamelet and PaSR cases respectively. As can be expected, the largest variation in the model usage for both cases is observed in and around the flame itself. A point of difference between the two cases that is evident from the figure is that the PPAC-ISAT run using models generated from the PaSR database handles the compositions in the unmixed pilot using a model with a noticeably smaller number of reactions compared to the run using reduced models generated with the flamelet database. Specifically, for the PaSR run these compositions are resolved with a model (M7) containing only $18\%$ of the reactions in the detailed mechanism, while for the PPAC-ISAT run using the flamelet database these compositions are resolved using a model (M4) that contains $59\%$ of the reactions in the detailed mechanism. 

\begin{figure}
	\centering
	\subfloat{%
		\resizebox*{0.3125\textwidth}{!}{\includegraphics{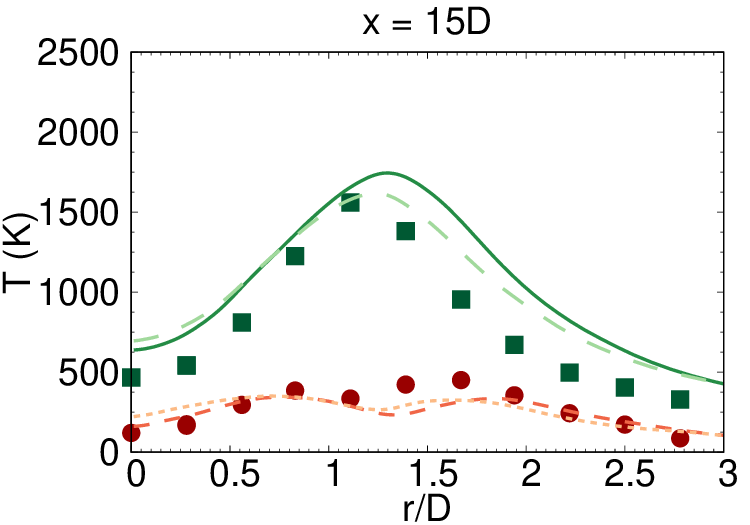}}}
	\subfloat{%
		\resizebox*{0.3125\textwidth}{!}{\includegraphics{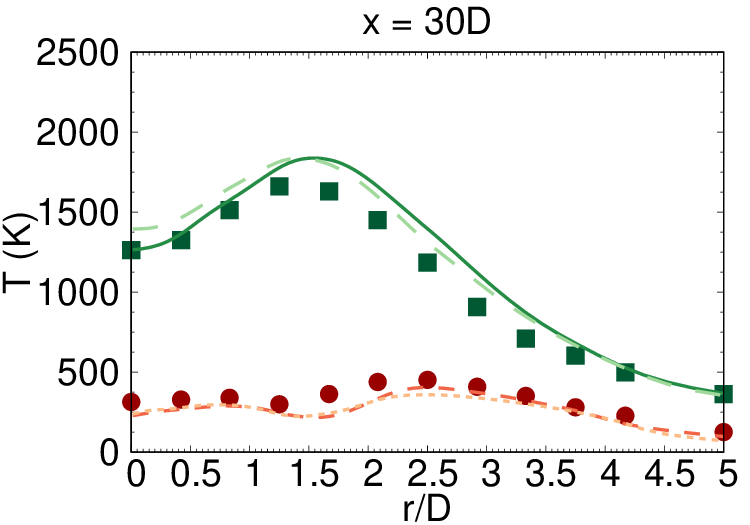}}}
	\subfloat{%
		\resizebox*{0.3125\textwidth}{!}{\includegraphics{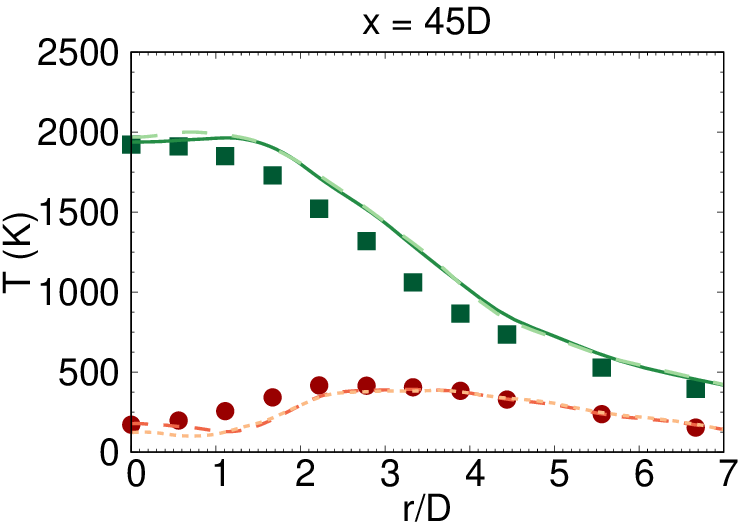}}}\\
	\subfloat{%
		\resizebox*{0.3125\textwidth}{!}{\includegraphics{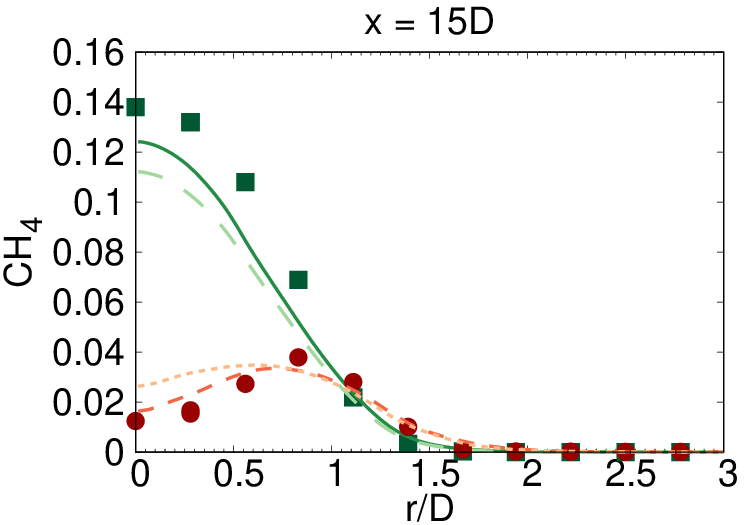}}}
	\subfloat{%
		\resizebox*{0.3125\textwidth}{!}{\includegraphics{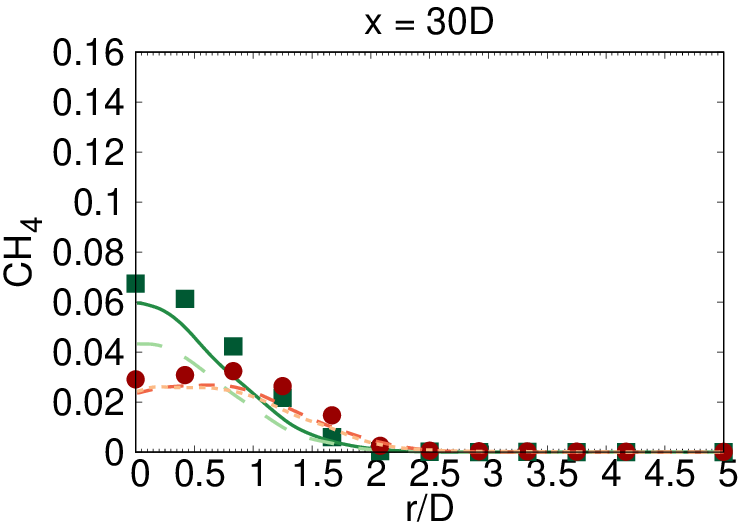}}}
	\subfloat{%
		\resizebox*{0.3125\textwidth}{!}{\includegraphics{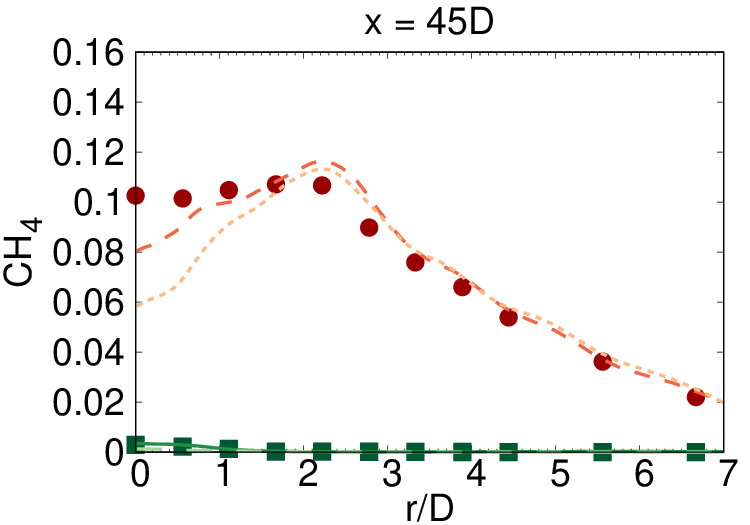}}}\\
		\subfloat{%
		\resizebox*{0.3125\textwidth}{!}{\includegraphics{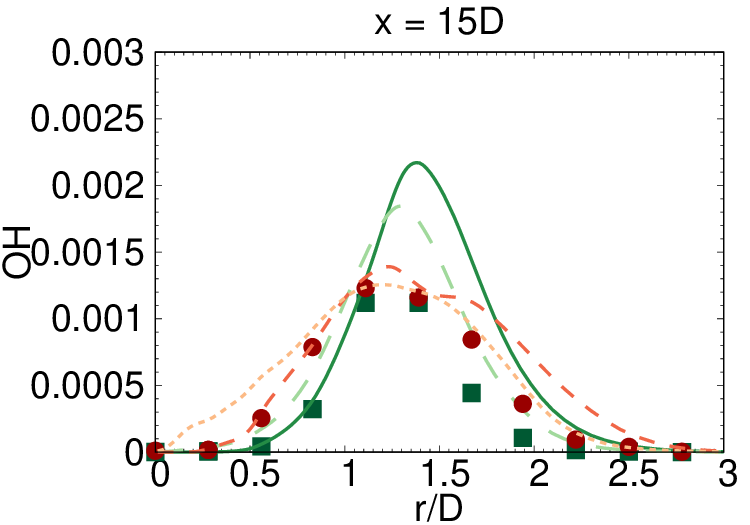}}}
	\subfloat{%
		\resizebox*{0.3125\textwidth}{!}{\includegraphics{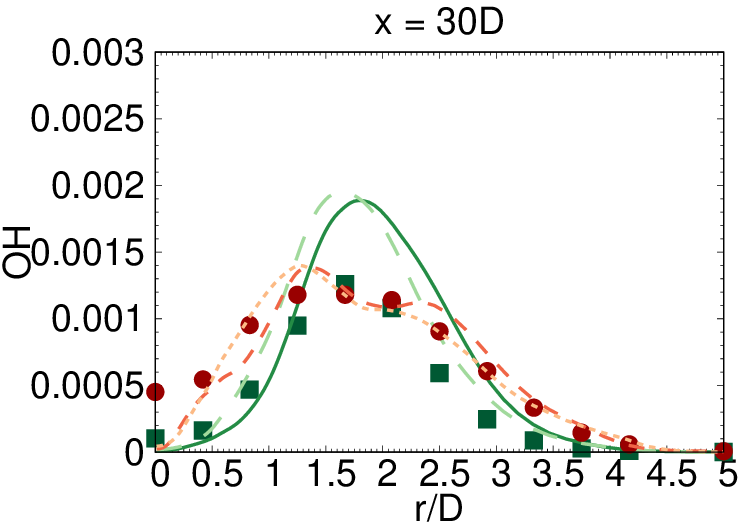}}}
	\subfloat{%
		\resizebox*{0.3125\textwidth}{!}{\includegraphics{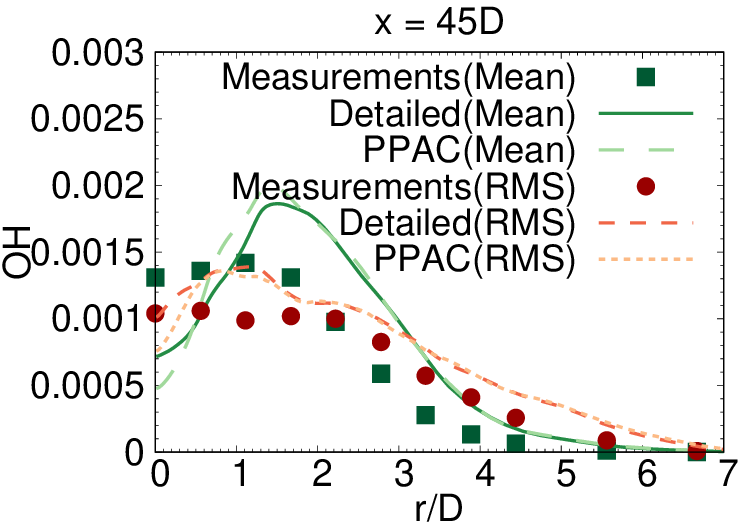}}}
	\caption{Comparison of measured and computed resolved mean and RMS of temperature and select species mass fractions. The computed values are shown for LES/PDF using ISAT with the detailed mechanism and LES/PDF using PPAC-ISAT with the flamelet database. The statistics are collected over two flow-through times.}
	\label{fig_flamelet}
\end{figure}

\begin{figure}
	\centering
	\subfloat{%
		\resizebox*{0.3125\textwidth}{!}{\includegraphics{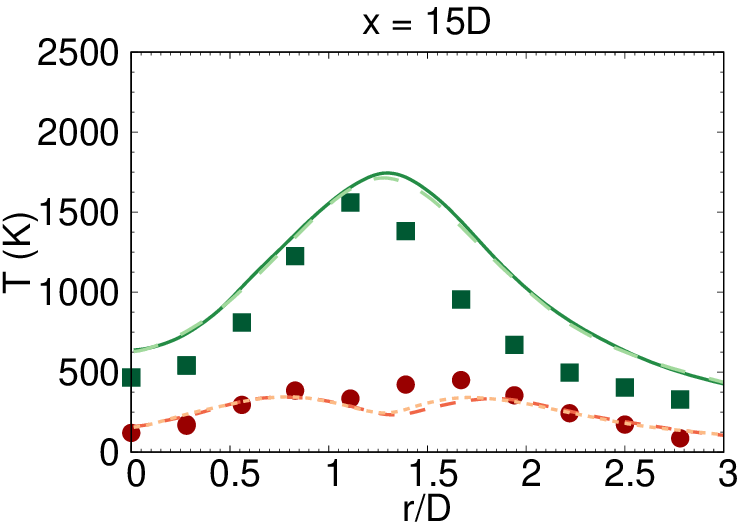}}}
	\subfloat{%
		\resizebox*{0.3125\textwidth}{!}{\includegraphics{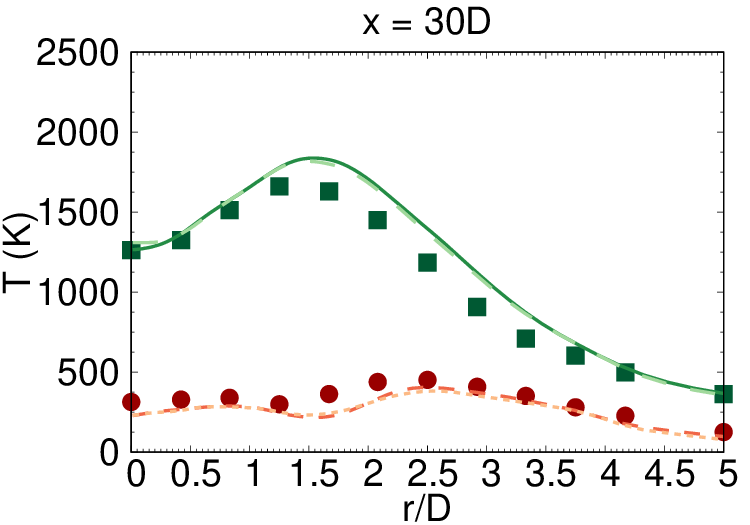}}}
	\subfloat{%
		\resizebox*{0.3125\textwidth}{!}{\includegraphics{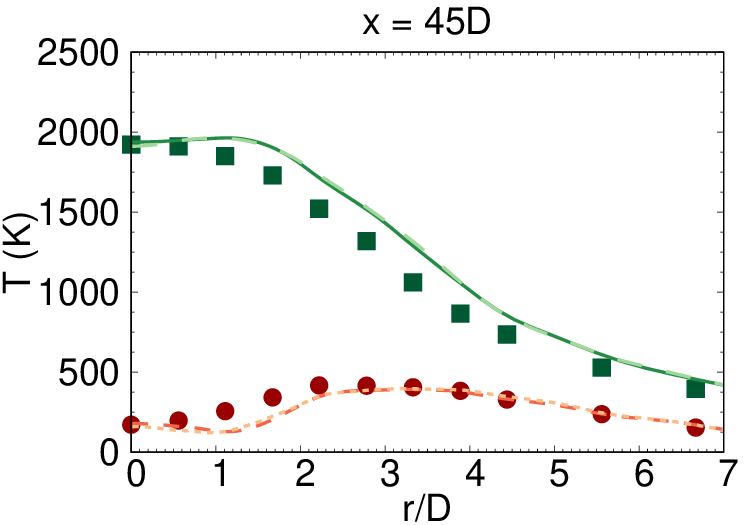}}}\\
	\subfloat{%
		\resizebox*{0.3125\textwidth}{!}{\includegraphics{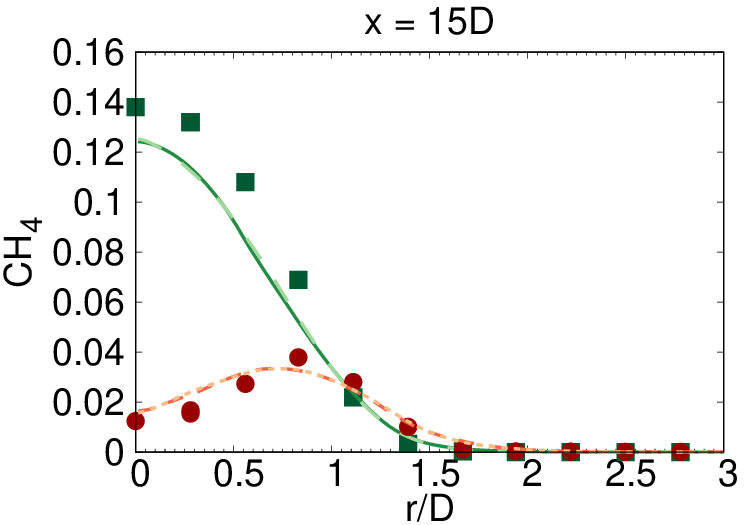}}}
	\subfloat{%
		\resizebox*{0.3125\textwidth}{!}{\includegraphics{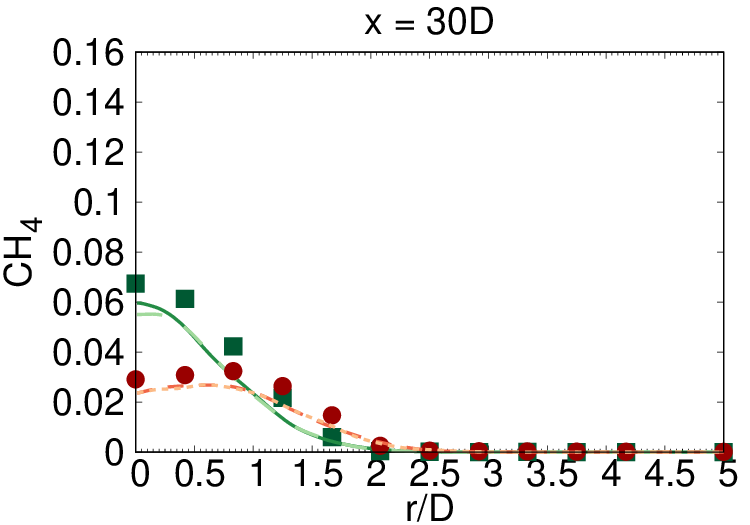}}}
	\subfloat{%
		\resizebox*{0.3125\textwidth}{!}{\includegraphics{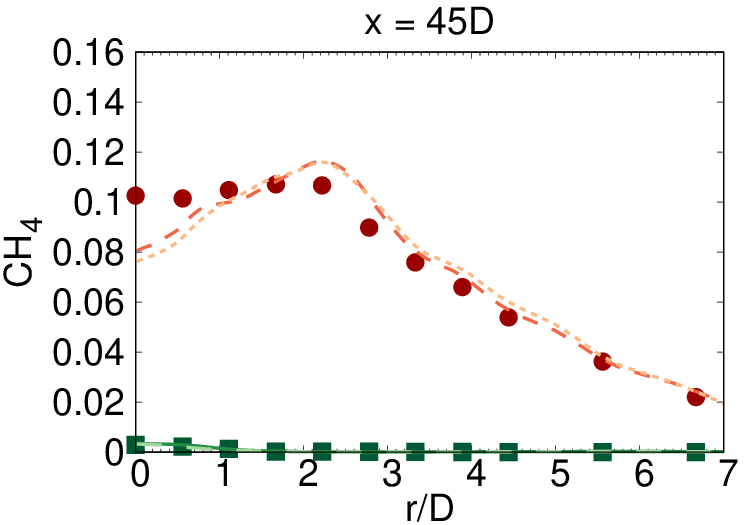}}}\\
		\subfloat{%
		\resizebox*{0.3125\textwidth}{!}{\includegraphics{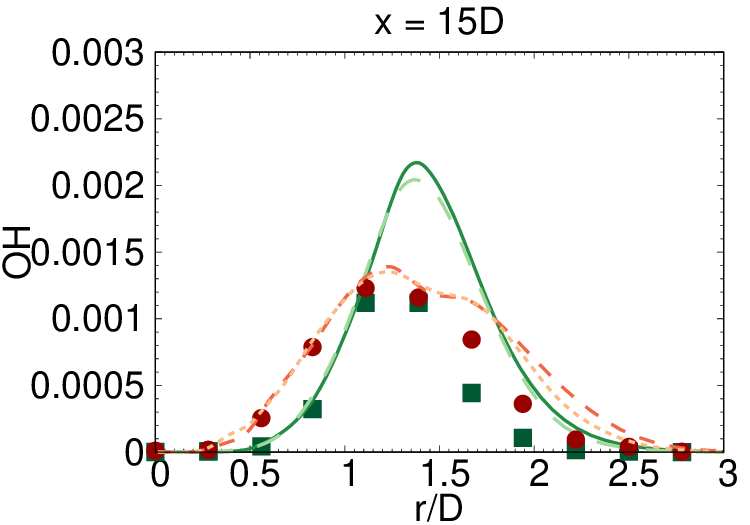}}}
	\subfloat{%
		\resizebox*{0.3125\textwidth}{!}{\includegraphics{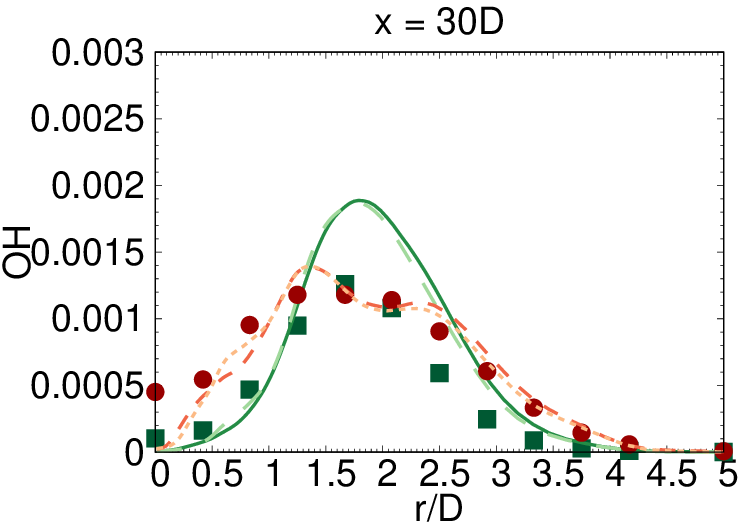}}}
	\subfloat{%
		\resizebox*{0.3125\textwidth}{!}{\includegraphics{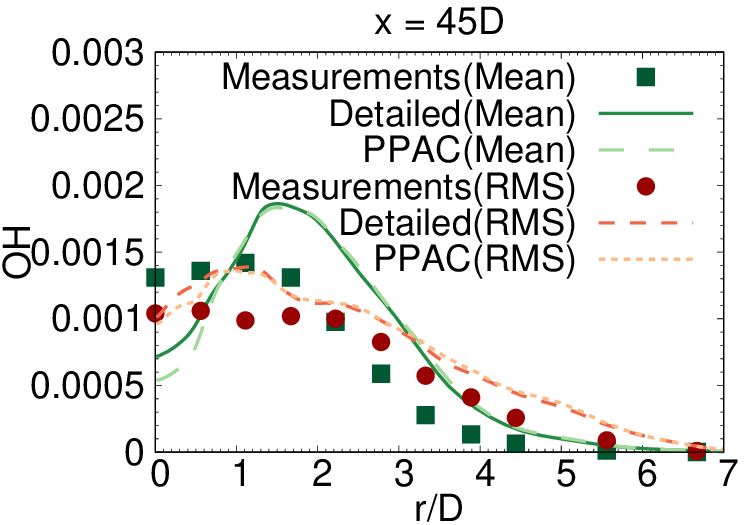}}}
	\caption{Comparison of measured and computed resolved mean and RMS of temperature and select species mass fractions. The computed values are shown for LES/PDF using ISAT with the detailed mechanism and LES/PDF using PPAC-ISAT with the PaSR database. The statistics are collected over two flow-through times.}
	\label{fig_pasr}
\end{figure}

A comparison of the radial profiles of the resolved mean and RMS of temperature, CH$_4$, and OH for the simulation using ISAT with the detailed mechanism, PPAC-ISAT, and the experimental measurements are shown in Figs. \ref{fig_flamelet} and \ref{fig_pasr}. The results for the PPAC-ISAT simulation using the flamelet database are shown in Fig. \ref{fig_flamelet}, and for the simulation using the PaSR database are shown in Fig. \ref{fig_pasr}. The statistics are collected over two flow-through times based on the mean jet inflow velocity. We observe that there is reasonable agreement between the PPAC-ISAT simulation with the flamelet database and the simulation using ISAT with the detailed mechanism at locations farther downstream of the jet exit ($x/D=30$, and $x/D=45$). However, there are noticeable discrepancies between results of the two simulations at the axial location closest to the jet exit ($x/D=15$). Figure \ref{fig_pasr} shows that excellent agreement with the simulation using ISAT with the detailed mechanism is attained by the simulation using PPAC-ISAT with the PaSR database for all the downstream locations explored here.



For a more quantitative assessment of the difference between the PPAC-ISAT simulations using the two different databases and the simulation using ISAT with the detailed mechanism, we use the normalized root-mean-square-difference (RMSD) metric proposed by Hiremath et al. \cite{hiremath2013large}. This metric is computed as follows:
\begin{equation}
    \epsilon(\xi) = \dfrac{[\xi^r-\xi^f]_{rms}}{\xi_{ref}},
\end{equation}
where $\xi^r$ is the value of the statistic obtained from the simulation using PPAC-ISAT, $\xi^f$ is the corresponding value obtained from the simulation using ISAT with the detailed mechanism. $\xi_{ref}$ is a reference quantity used for normalization, which is taken to be the maximum value of the resolved mean over radial profiles at all axial locations for species mass fractions and $1000$K for temperature. The metric is computed using data for radial profiles of the resolved mean and RMS at all the downstream locations explored here. The values of this metric for select species mass fractions and temperature is shown in Tab. \ref{tab_combined} for the PPAC-ISAT simulations using the flamelet and PaSR databases. The quantitative results shown in these tables are in line with the qualitative observations made previously. Overall, we observe that the PPAC-ISAT flamelet simulation matches the resolved mean and RMS of the simulation using ISAT with the detailed mechanism to within $7\%$ for the major species mass fractions and temperature and OH species mass fraction profiles to within $8.5\%$. The PPAC-ISAT simulation using the PaSR database matches the statistics of the simulation using ISAT with the detailed mechanism to within approximately $2\%$ for all the species mass fractions and temperature. \par

\begin{table}
	\centering
	\caption{Normalized root mean square differences between LES/PDF using ISAT with the detailed mechanism and PPAC-ISAT using the flamelet and PaSR databases in the offline preprocessing stage}\label{tab_combined}
	\begin{tabular}{ccccc}
	    \hline
	    &\multicolumn{2}{c}{\bfseries Flamelet} & \multicolumn{2}{c}{\bfseries PaSR}\\
		\hline
		\multirow{2}{*}{Quantity ($\xi$)}&   \multicolumn{1}{p{2.1 cm}}{\centering RMSD in \\ $\langle \xi \rangle$ ($\%$)}
		 & \multicolumn{1}{p{2.1 cm}}{\centering RMSD in\\ $\xi_{rms}$ ($\%$)}& \multicolumn{1}{p{2.1 cm}}{\centering RMSD in\\ $\langle \xi \rangle$ ($\%$)} & \multicolumn{1}{p{2.1 cm}}{\centering RMSD in\\ $\xi_{rms}$ ($\%$)}\\
		\hline
		T & 6.4 & 2.7 & 1.5 & 1.3 \\
		CH$_4$&  4.1 & 2.3 & 0.6 & 0.8 \\
		O$_2$&   3.6 & 1.4 & 0.9 & 0.7 \\
		CO$_2$&  4.2 &  1.9 &  1.0 &  0.8 \\
		OH &  8.4 &  5.5 &  1.3 &  2.1 \\\hline
	\end{tabular}
\end{table}

The PPAC-ISAT simulations using the flamelet and PaSR databases lead to a reduction in the average wall clock time per time step over a simulation using the detailed mechanism with ISAT by $33\%$ and $38\%$ respectively. This is comparable to the reduction in the average wall clock time per time step ($39\%$) obtained with PPAC-ISAT using a database of compositions extracted directly from a LES/PDF simulation of the Cartesian configuration using the detailed mechanism \cite{newale2020comp}. This shows that it is indeed feasible to generate the initial database efficiently using low-dimensional simulations and attain a comparable performance to the optimal case, in which directly relevant compositions were extracted from a detailed LES/PDF simulation. The feasibility of database generation using low-dimensional simulations obviates the need for having access to detailed turbulent combustion simulations for using PPAC. This not only represents a significant computational saving, since a detailed LES/PDF simulation is replaced by a series of low-dimensional simulations, whose cost is negligible, but it also provides confidence that this adaptive framework can be used for cases where the detailed LES/PDF simulation would be prohibitively expensive. 



\begin{figure}
    \centering
    \subfloat[Original flamelet database]{%
		\resizebox*{0.3125\textwidth}{!}{\includegraphics{rt_flamelet_loc_test/combined_T_xd_15.eps}}
		\resizebox*{0.3125\textwidth}{!}{\includegraphics{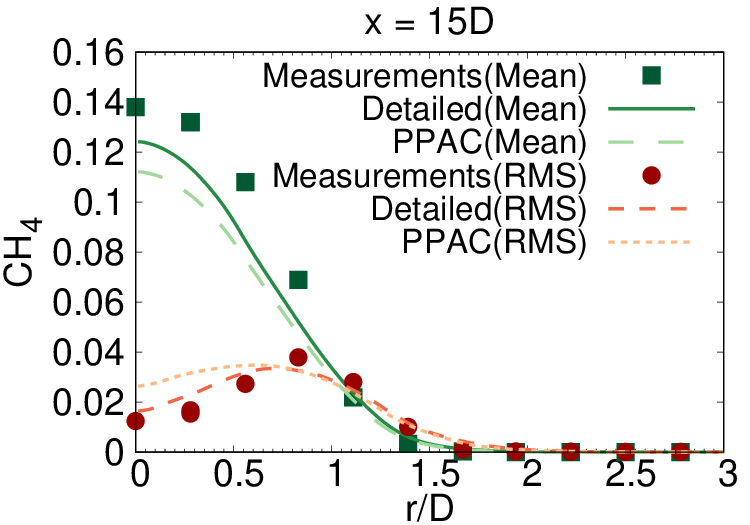}}
		\resizebox*{0.3125\textwidth}{!}{\includegraphics{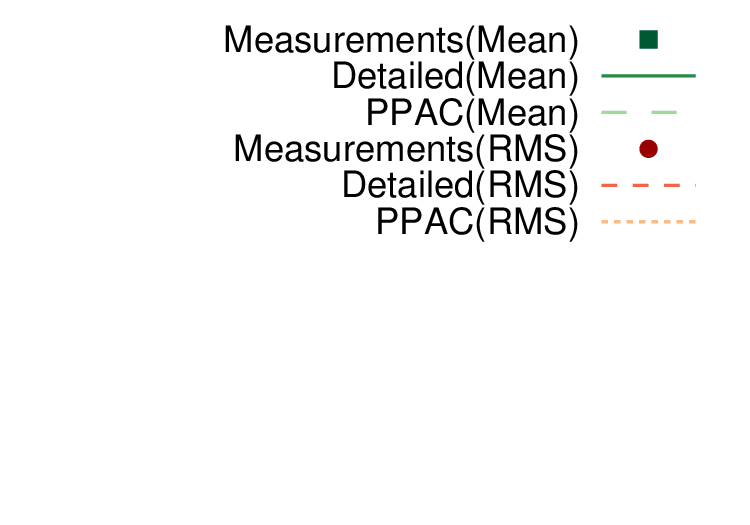}}}\\
    \subfloat[Partial online flamelet database]{%
		\resizebox*{0.3125\textwidth}{!}{\includegraphics{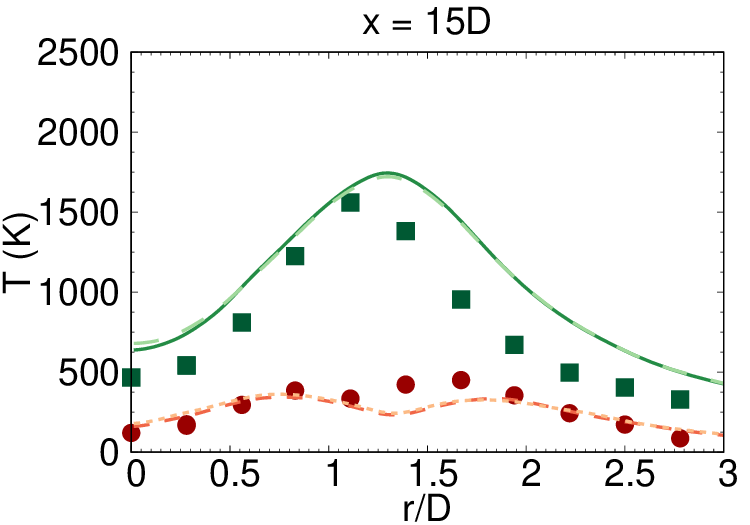}}
		\resizebox*{0.3125\textwidth}{!}{\includegraphics{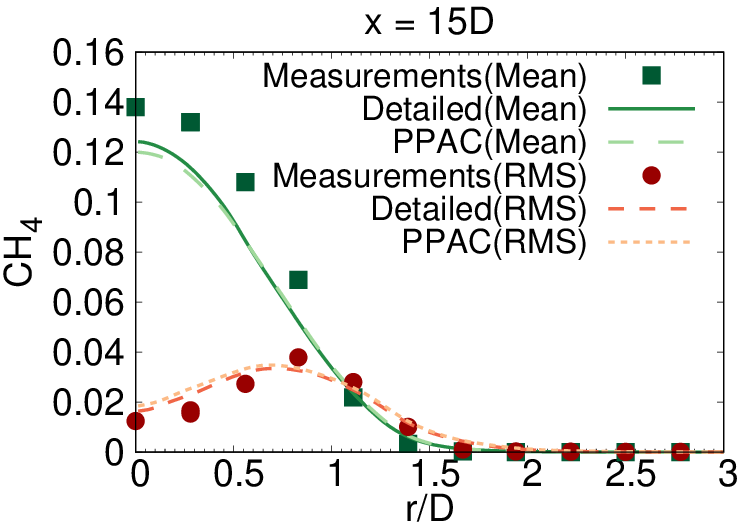}}
		\resizebox*{0.3125\textwidth}{!}{\includegraphics{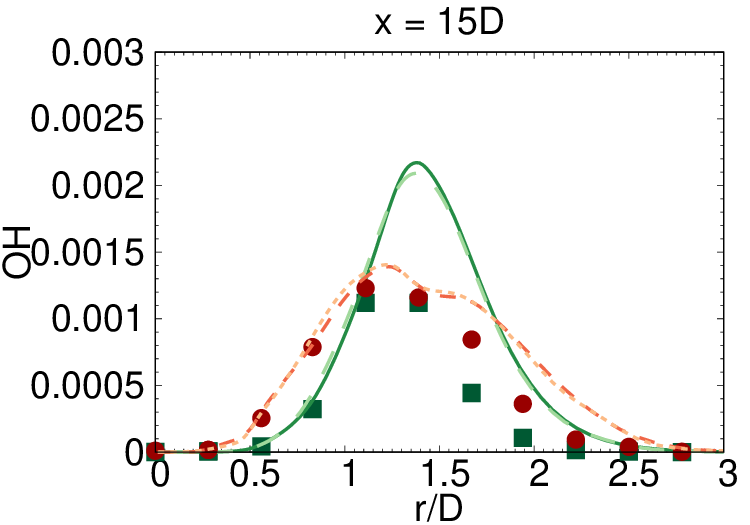}}}
    \caption{Comparison of measured and computed resolved mean and RMS of temperature and select species mass fractions at $15D$ downstream of the jet exit. The top row shows the computed values for LES/PDF using ISAT with the detailed mechanism and LES/PDF using PPAC-ISAT with the flamelet database for reference. The bottom row shows the LES/PDF using ISAT with the detailed mechanism and LES/PDF using PPAC-ISAT with region-specific reduced mechanisms for problematic regions derived again using online compositions and partitioning based on the initial database generated using flamelets. The statistics are collected over two flow-through times.}
	\label{fig_ponline}
\end{figure}

\begin{table}
	\centering
	\caption{Normalized root mean square differences between LES/PDF using ISAT with the detailed mechanism and PPAC-ISAT using the original flamelet database and a modified database using online compositions for problematic regions in the offline preprocessing stage}\label{tab_compare_orig_up}
	\begin{tabular}{ccccc}
	    \hline
	    &\multicolumn{2}{c}{\bfseries Flamelet (Original)} & \multicolumn{2}{c}{\bfseries Flamelet (Partial online)}\\
		\hline
		\multirow{2}{*}{Quantity ($\xi$)}&   \multicolumn{1}{p{2.1 cm}}{\centering RMSD in \\ $\langle \xi \rangle$ ($\%$)}
		 & \multicolumn{1}{p{2.1 cm}}{\centering RMSD in\\ $\xi_{rms}$ ($\%$)}& \multicolumn{1}{p{2.1 cm}}{\centering RMSD in\\ $\langle \xi \rangle$ ($\%$)} & \multicolumn{1}{p{2.1 cm}}{\centering RMSD in\\ $\xi_{rms}$ ($\%$)}\\
		\hline
		T & 6.4 & 2.7 & 2.2 & 1.2 \\
		CH$_4$&  4.1 & 2.3 & 1.3 & 0.8 \\
		O$_2$&   3.6 & 1.4 & 1.1 & 0.6 \\
		CO$_2$&  4.2 &  1.9 &  1.7 & 0.8  \\
		OH &  8.4 &  5.5 & 3.6  & 1.7  \\\hline
	\end{tabular}
\end{table}
To determine the reasons for the relatively higher discrepancies between the PPAC-ISAT simulation using the flamelet database and the simulation using ISAT with the detailed mechanism, we examine the region-specific reaction mapping errors incurred due to the use PPAC-ISAT in the LES/PDF simulation. The regions where the reaction mapping errors exceed the reduction threshold are deemed to be problematic. Here, two of the ten regions fall into this category. We hypothesize that this situation arises because the database of compositions used to derive the reduced kinetic models for these two regions was not representative of the actual compositions encountered at runtime, and classified into those regions. To test this claim, a set of reduced mechanisms for only these specific regions were derived again with DRGEP \cite{pepiot2008efficient} using $500$ compositions extracted from the statistically stationary LES/PDF simulation using ISAT with the detailed mechanism that are classified as belonging to these regions. A reduced mechanism is then selected for each of these problematic regions from the corresponding set, such that it satisfies the reduction error threshold of $10^{-5}$, and contains the smallest number of species. The radial profiles for temperature, CH$_4$ and OH at $15$D downstream of the jet exit for the PPAC-ISAT simulation using the mixed set of models are shown in Fig. \ref{fig_ponline} in the bottom row, along with the corresponding results for the PPAC-ISAT simulation using the unmodified flamelet database in the top row. As the reduced models for two of the ten regions are generated using online compositions, we refer to this simulation as using a partial online flamelet database. The results for the PPAC-ISAT simulation using the original or unmodified flamelet database at $15$D that were shown in Fig. \ref{fig_flamelet}, are repeated for ease of comparison. For brevity, only the radial profiles at the axial location closest to the jet exit are shown as the largest discrepancy with reference simulation using ISAT with the detailed mechanism was observed at this location. A complete set of radial profiles for the PPAC-ISAT simulation using the mixed set of models is provided in appendix \ref{sec:app_a}. Comparing the top and bottom rows in Fig. \ref{fig_ponline}, we observe a noticeable improvement in the agreement with the simulation using ISAT with the detailed mechanism. The values of the RMSD metric for resolved mean and RMS of temperature and select species mass fractions, obtained for PPAC-ISAT using models generated with the original flamelet database, and PPAC-ISAT using the mixed set of models are shown in Tab. \ref{tab_compare_orig_up}. The improvement in the agreement is quantitatively confirmed with this metric. The regions identified as problematic contain particle compositions located in the coflow and locations where the coflow mixes with the pilot and further downstream where the coflow mixes with the products. This is likely caused by the fact that there is no analogue for the pilot stream in the flamelet configuration used in this work. 

To address this possible deficiency and ascertain this as the cause for the observed discrepancy, we augment the flamelet database with samples from two additional 1D counterflow flame configurations. Specifically, compositions are generated from two additional flamelet configurations, where the stream compositions are set to that of fuel-pilot and pilot-coflow. As before, both configurations are run in mixture fraction space to extinction. The compositions from the original 1D counterflow flame configuration using fuel-coflow compositions for the two streams are combined with compositions extracted from these two configurations to get the complete database. We subsequently refer to this combination as the augmented flamelet database.   

\begin{figure}
 	\centering
 		\resizebox*{0.4875\textwidth}{!}{\includegraphics{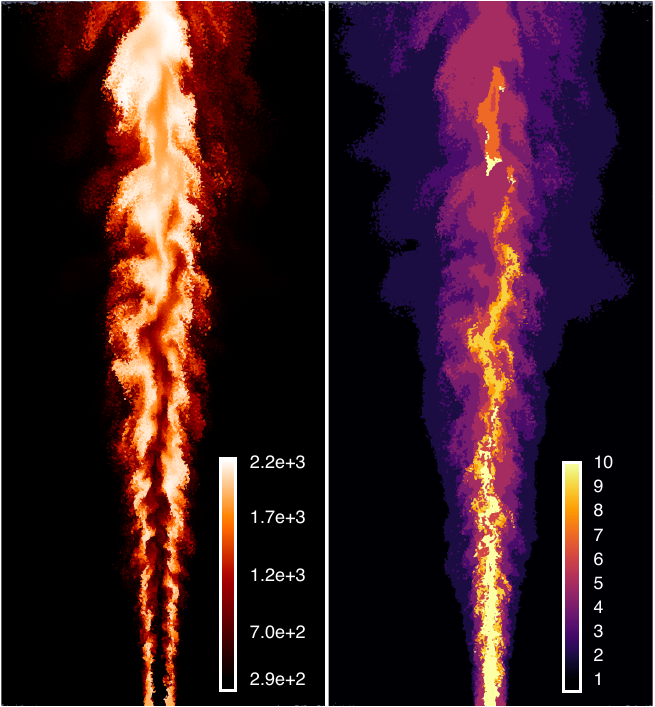}}
 		\resizebox*{0.4875\textwidth}{!}{\includegraphics[trim=0 0.5cm 0 0, clip]{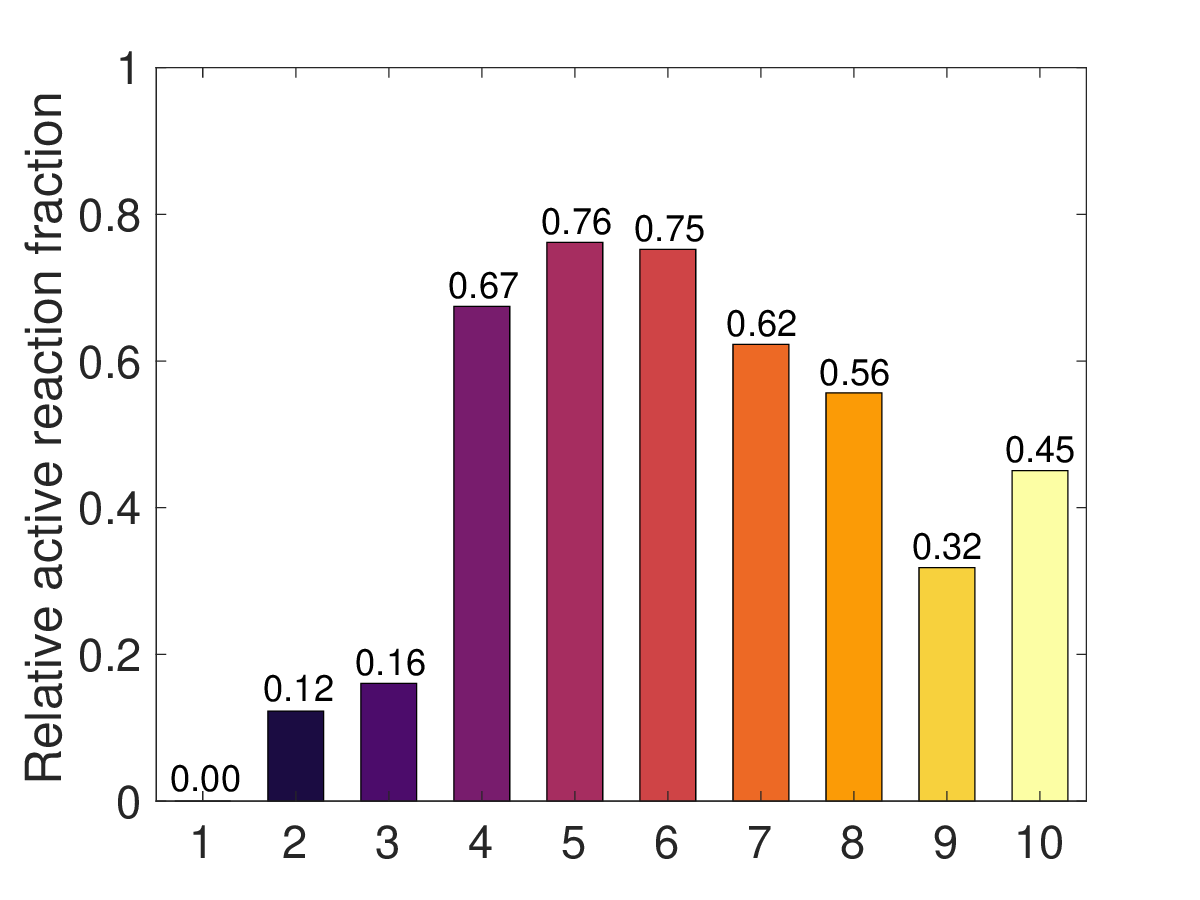}}
 	\caption{A cross-sectional view of the instantaneous PDF particle distribution colored by temperature (K) and model ID (left), and a histogram for the relative active reaction fraction for each region-specific reduced model (right). The results for the full-scale PPAC-ISAT LES/PDF simulation using reduced models generated with the augmented flamelet database.}
 	\label{fig_qualitative_aug_flamelet}
 \end{figure}

Figure \ref{fig_qualitative_aug_flamelet} shows a cross-sectional view of the instantaneous PDF particle temperature distribution along with corresponding model IDs on the left for the full-scale LES/PDF PPAC-ISAT run using this augmented flamelet database. The right half of the figure shows the relative active reaction fraction for the region-specific reduced models generated using the augmented flamelet database. As before, the color used for the bars in the histogram matches those used for the model IDs in the cross-sectional view for ease of understanding. We observe that the compositions in the unmixed coflow are handled using a reduced model with zero reactions. Additionally, the model utilized in the potential core of the fuel jet contains only $45\%$ of the reactions contained in the detailed model. As was observed for the original flamelet and PaSR database based runs, the largest variation in the model usage is observed close to the flame location. 

\begin{figure}
    \centering
   \subfloat[Original flamelet database]{%
		\resizebox*{0.3125\textwidth}{!}{\includegraphics{rt_flamelet_loc_test/combined_T_xd_15.eps}}
		\resizebox*{0.3125\textwidth}{!}{\includegraphics{combined_CH4_xd_15_Flamelet.eps}}
		\resizebox*{0.3125\textwidth}{!}{\includegraphics{rt_flamelet_loc_test/combined_OH_xd_15.eps}}}\\
		\subfloat[Augmented flamelet database]{%
                \resizebox*{0.3125\textwidth}{!}{\includegraphics{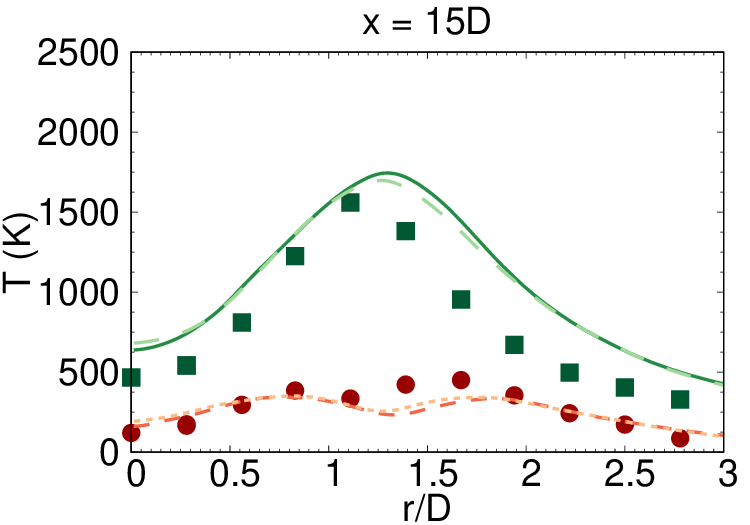}}
                \resizebox*{0.3125\textwidth}{!}{\includegraphics{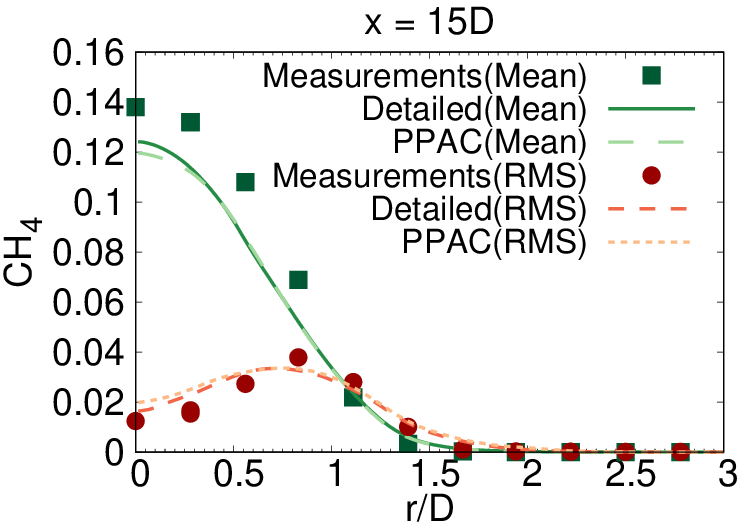}}
                \resizebox*{0.3125\textwidth}{!}{\includegraphics{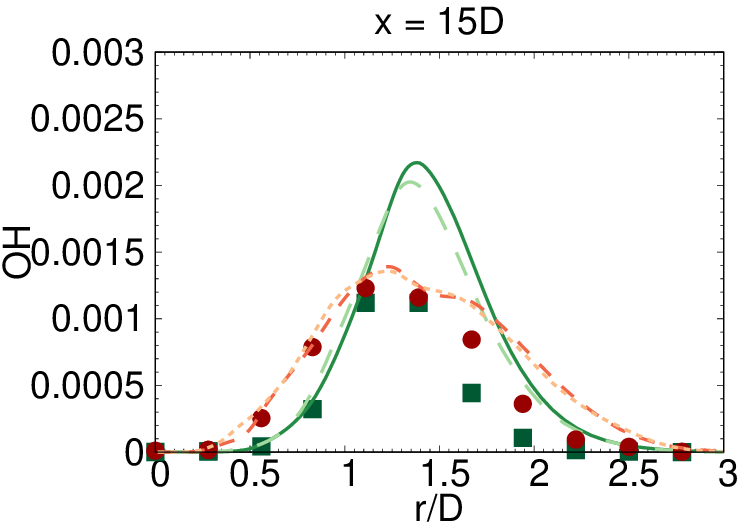}}}
     \caption{Comparison of measured and computed resolved mean and RMS of temperature and select species mass fractions at $15D$ downstream of the jet exit. The top row shows the computed values for LES/PDF using ISAT with the detailed mechanism and LES/PDF using PPAC-ISAT with the flamelet database for reference. The bottom row shows the LES/PDF using ISAT with the detailed mechanism and LES/PDF using PPAC-ISAT with the augmented flamelet database. The statistics are collected over two flow-through times.}
        \label{fig_augmentedFlamelet}
\end{figure}

Figure \ref{fig_augmentedFlamelet} shows the radial profiles for the resolved mean and RMS of temperature and select species mass fractions at $15$ diameters downstream of the jet exit, for the full-scale LES/PDF PPAC-ISAT simulation using reduced models generated with the augmented flamelet database in the bottom row and for ease of comparison, the PPAC-ISAT simulation using the original flamelet database in the top row. As before, we only show results for the location closest to the jet exit, with the complete set of radial profiles being provided in the appendix \ref{sec:app_a}. Comparing the two sets of results shown in Fig. \ref{fig_augmentedFlamelet}, we observe noticeably improved agreement at this axial location between the PPAC-ISAT simulation using the augmented flamelet database and the reference simulation using ISAT with the detailed mechanism. This exercise further supports the claim that the accounting for the pilot is crucial for creating an appropriate database for these types of piloted flames. Table \ref{tab_compare_orig_augmented} shows the normalized RMSD values between the full-scale PPAC-ISAT simulation using the augmented flamelet database and the simulation using ISAT with the detailed mechanism. For comparison, the corresponding RMSD values for the PPAC-ISAT simulation using the original flamelet database are shown on the left. These values provide quantitative confirmation of the improved agreement attained with the use of the augmented flamelet database with the reference simulation using ISAT with the detailed mechanism. For completeness, we also directly quantify the conservation and incurred errors in the Cartesian configuration for PPAC and PPAC-ISAT using reduced models generated with the augmented flamelet database, which are reported in the appendix \ref{sec:app_b}. 

\begin{table}
	\centering
	\caption{Normalized root mean square differences between LES/PDF using ISAT with the detailed mechanism and PPAC-ISAT using the original flamelet and augmented flamelet databases in the offline preprocessing stage}\label{tab_compare_orig_augmented}
	\begin{tabular}{ccccc}
	    \hline
	    &\multicolumn{2}{c}{\bfseries Flamelet (Original)} & \multicolumn{2}{c}{\bfseries Flamelet (Augmented)}\\
		\hline
		\multirow{2}{*}{Quantity ($\xi$)}&   \multicolumn{1}{p{2.1 cm}}{\centering RMSD in \\ $\langle \xi \rangle$ ($\%$)}
		 & \multicolumn{1}{p{2.1 cm}}{\centering RMSD in\\ $\xi_{rms}$ ($\%$)}& \multicolumn{1}{p{2.1 cm}}{\centering RMSD in\\ $\langle \xi \rangle$ ($\%$)} & \multicolumn{1}{p{2.1 cm}}{\centering RMSD in\\ $\xi_{rms}$ ($\%$)}\\
		\hline
		T & 6.4 & 2.7 & 2.2 & 2.0 \\
		CH$_4$&  4.1 & 2.3 & 0.8 & 1.5 \\
		O$_2$&   3.6 & 1.4 & 1.4 & 1.0 \\
		CO$_2$&  4.2 &  1.9 &  1.6 & 1.3  \\
		OH &  8.4 &  5.5 & 3.7  & 3.0  \\\hline
	\end{tabular}
\end{table}

\section{Summary and conclusions}
A key assumption in the PPAC framework is that compositions in the initial database used in the offline preprocessing stage are representative of those encountered at runtime. Consequently, the efficient generation of the initial database is crucial to the long term success of PPAC. In this work, we examine the suitability of the 1D counterflow flames and PaSR to generate the initial database. PPAC and PPAC-ISAT simulations using the flamelet and PaSR databases lead to acceptable conservation and incurred errors for the small-domain Cartesian configuration. For the full-scale cylindrical configuration, the PPAC-ISAT simulation using the PaSR database leads to excellent agreement with the reference simulation using ISAT with the detailed mechanism. A larger discrepancy is observed between the PPAC-ISAT simulation using the flamelet database and the reference simulation, which is investigated in detail. We show that accuracy comparable to that obtained with the PaSR database can be achieved by suitably augmenting the flamelet database to account for the presence of the pilot in flame D. 

 The PPAC-ISAT simulation using the PaSR database leads to a reduction in average wall clock time per time step of $38\%$ over a simulation using ISAT with the detailed mechanism for the full-scale configuration. This reduction is obtained while matching the resolved mean and RMS of species mass fraction and temperature from the simulation using ISAT with the detailed mechanism to within $2.2\%$. This performance is at the same level as the optimal case \cite{newale2020comp}, where the requirement of the database compositions being representative of those encountered at runtime is trivially satisfied. This shows that it is indeed feasible to utilize PPAC with initial databases generated using computationally-efficient low-dimensional simulations to significantly accelerate turbulent combustion simulations with minimal loss of accuracy.   
 

\section{Acknowledgements}
This research was funded by the US Department of Energy Office of Science, Office of Basic Energy Sciences under award number DE-FG02-90ER14128.

\section{References}


\bibliographystyle{tfq}
\bibliography{interacttfqsample}


\section{Appendices}

\appendix

\section{Full scale results: Partial online and augmented flamelet databases}
\label{sec:app_a}

\begin{figure}
        \centering
        \subfloat{%
                \resizebox*{0.3125\textwidth}{!}{\includegraphics{rt_spdf_online_compo_test/combined_T_xd_15.eps}}}
        \subfloat{%
                \resizebox*{0.3125\textwidth}{!}{\includegraphics{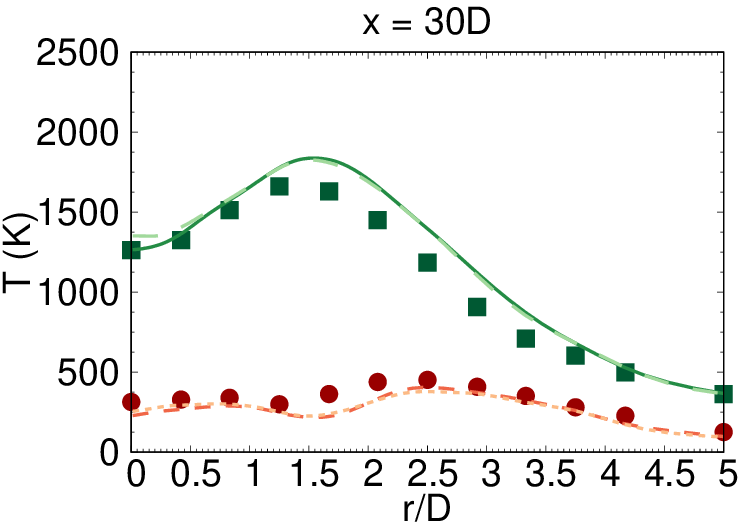}}}
        \subfloat{%
                \resizebox*{0.3125\textwidth}{!}{\includegraphics{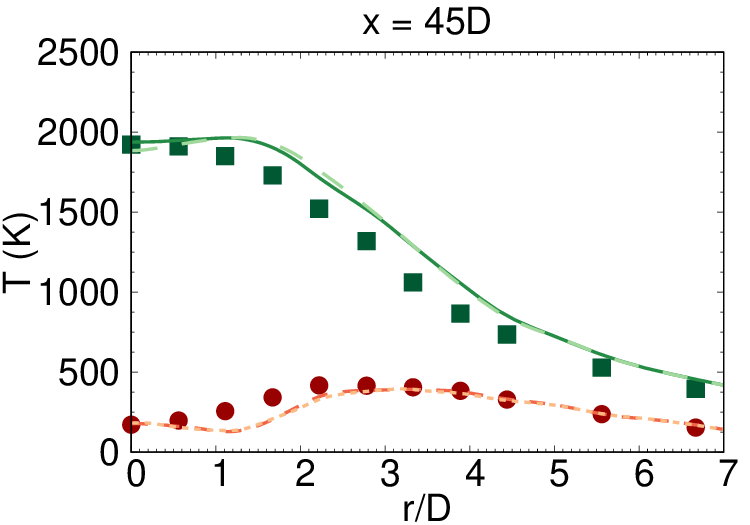}}}\\
        \subfloat{%
                \resizebox*{0.3125\textwidth}{!}{\includegraphics{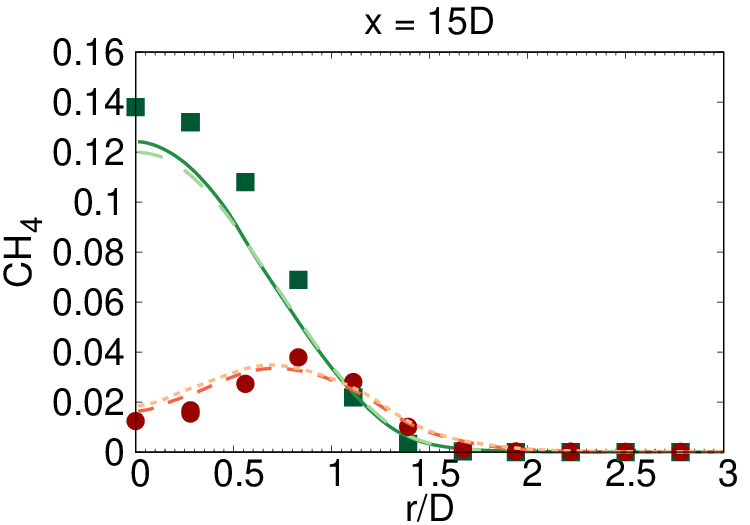}}}
        \subfloat{%
                \resizebox*{0.3125\textwidth}{!}{\includegraphics{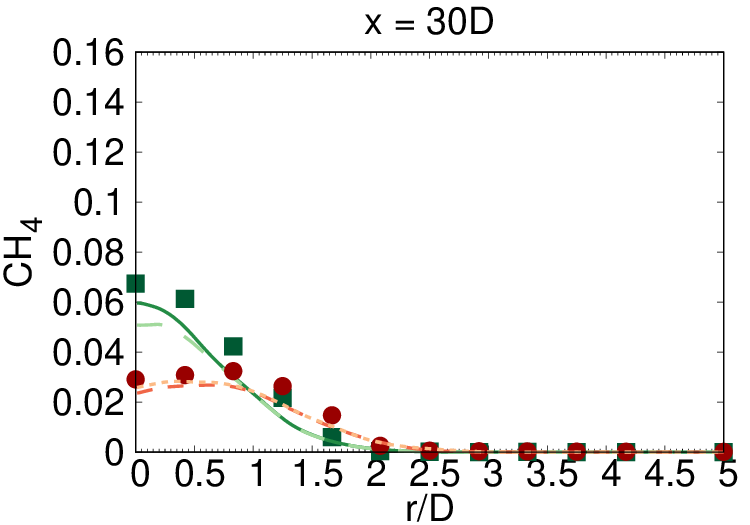}}}
        \subfloat{%
                \resizebox*{0.3125\textwidth}{!}{\includegraphics{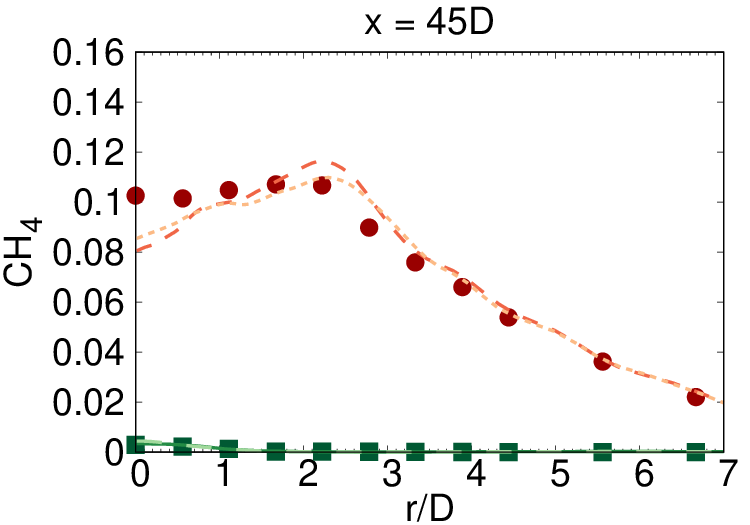}}}\\
                \subfloat{%
                \resizebox*{0.3125\textwidth}{!}{\includegraphics{rt_spdf_online_compo_test/combined_OH_xd_15.eps}}}
        \subfloat{%
                \resizebox*{0.3125\textwidth}{!}{\includegraphics{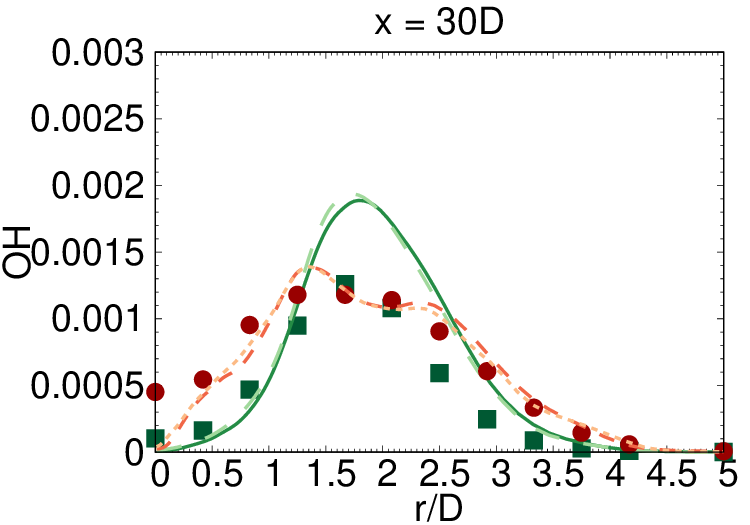}}}
        \subfloat{%
                \resizebox*{0.3125\textwidth}{!}{\includegraphics{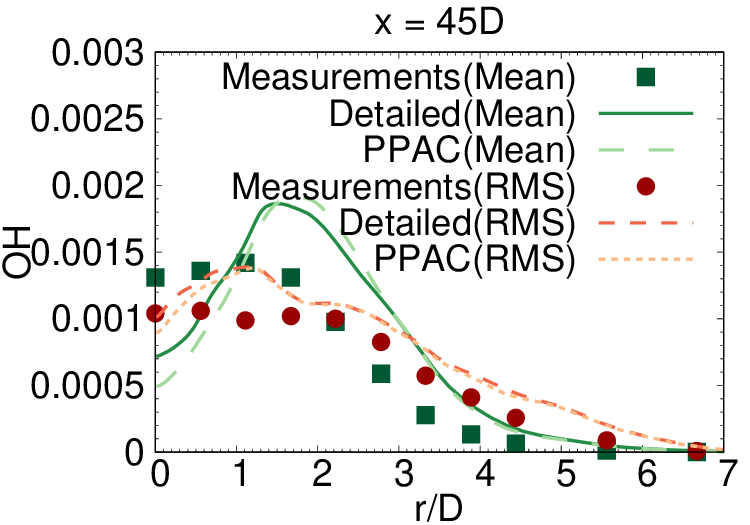}}}
        \caption{Comparison of measured and computed resolved mean and RMS of temperature and select species mass fractions. The computed values are shown for LES/PDF using ISAT with the detailed mechanism and LES/PDF using PPAC-ISAT with the region-specific reduced mechanisms for problematic regions derived again using online compositions and partitioning based on the initial database generated using flamelets. The statistics are collected over two flow-through times.}
        \label{fig_ponlineAppendix}
 \end{figure}

\begin{figure}
         \centering
         \subfloat{%
                 \resizebox*{0.3125\textwidth}{!}{\includegraphics{rt_augmented_flamelet_loc_test/combined_T_xd_15.eps}}}
         \subfloat{%
                 \resizebox*{0.3125\textwidth}{!}{\includegraphics{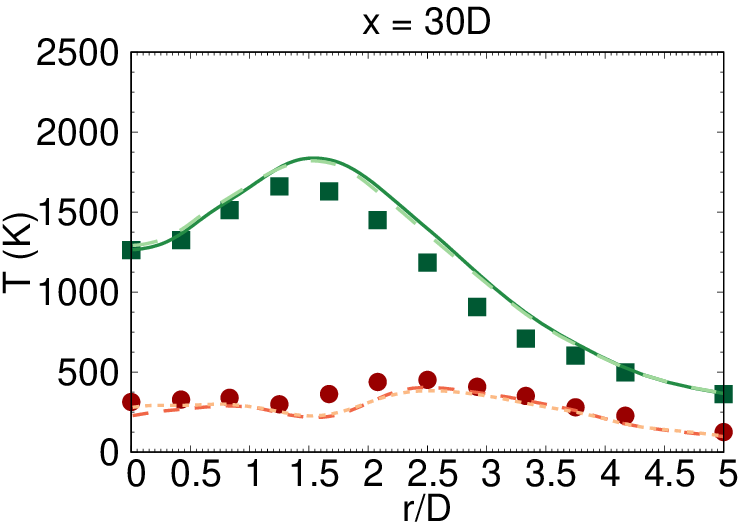}}}
         \subfloat{%
                 \resizebox*{0.3125\textwidth}{!}{\includegraphics{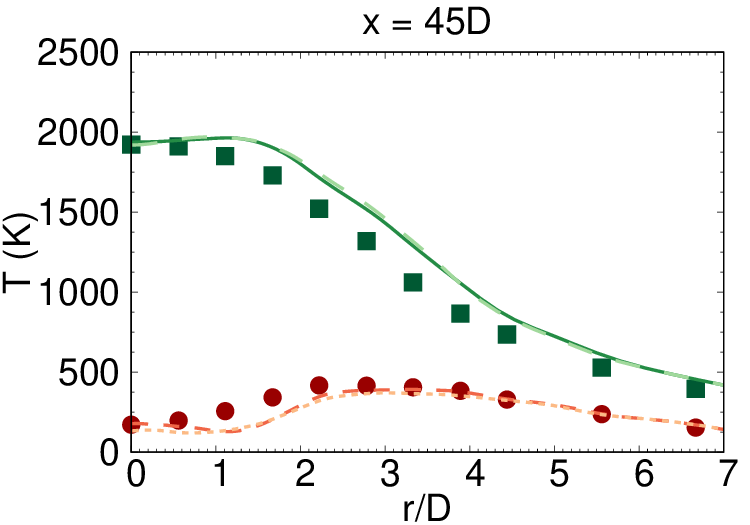}}}\\
         \subfloat{%
                 \resizebox*{0.3125\textwidth}{!}{\includegraphics{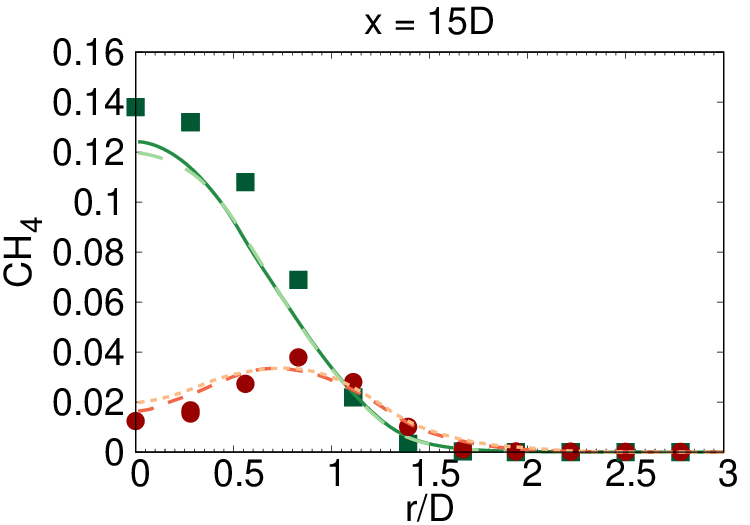}}}
         \subfloat{%
                 \resizebox*{0.3125\textwidth}{!}{\includegraphics{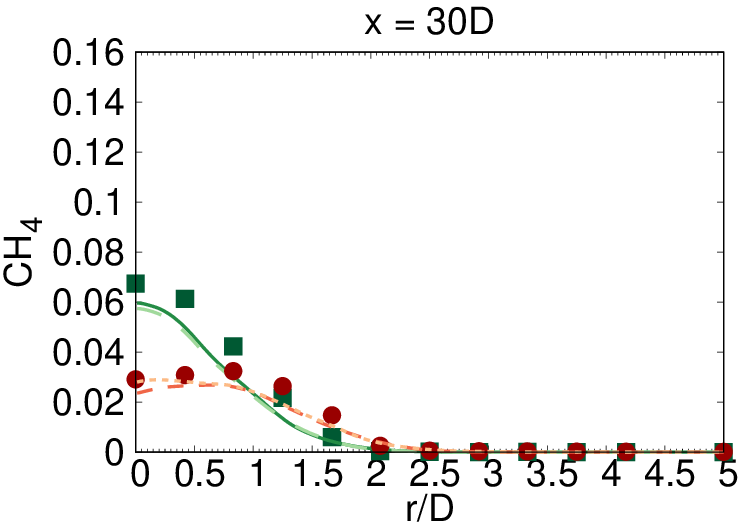}}}
         \subfloat{%
                 \resizebox*{0.3125\textwidth}{!}{\includegraphics{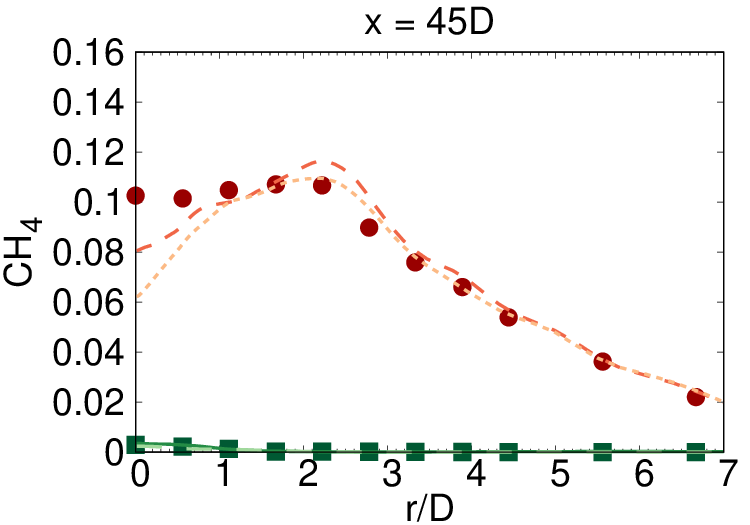}}}\\
                 \subfloat{%
                 \resizebox*{0.3125\textwidth}{!}{\includegraphics{rt_augmented_flamelet_loc_test/combined_OH_xd_15_pAug.eps}}}
         \subfloat{%
                 \resizebox*{0.3125\textwidth}{!}{\includegraphics{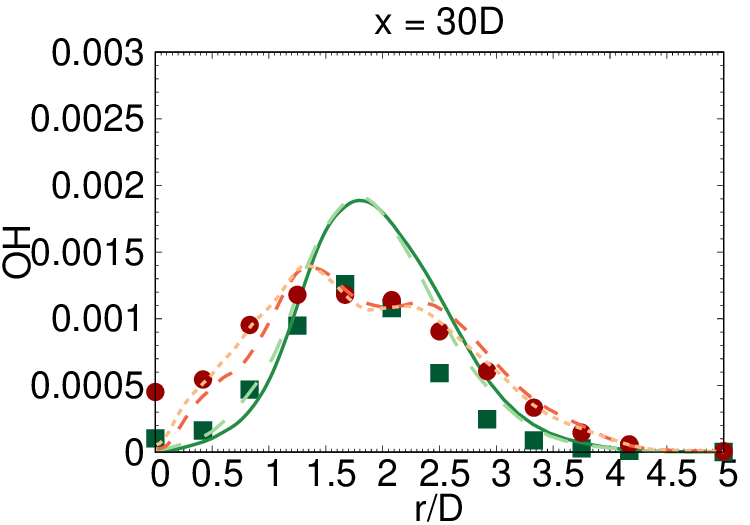}}}
         \subfloat{%
                 \resizebox*{0.3125\textwidth}{!}{\includegraphics{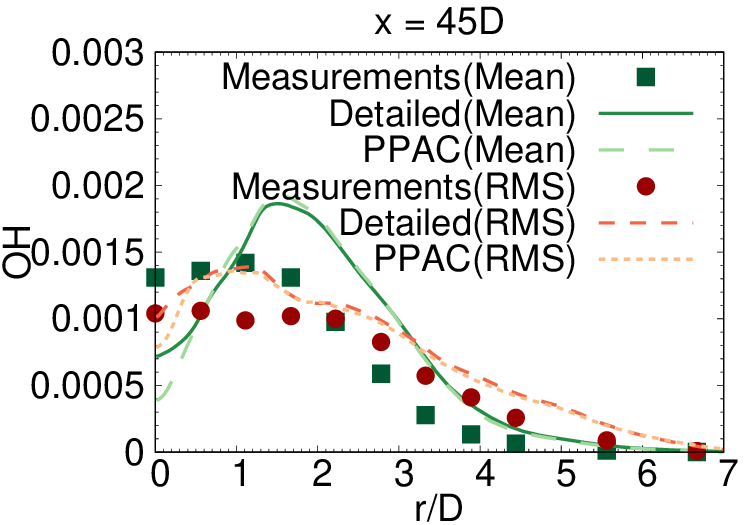}}}
         \caption{Comparison of measured and computed resolved mean and RMS of temperature and select species mass fractions. The computed values are shown for LES/PDF using ISAT with the detailed mechanism and LES/PDF using PPAC-ISAT with the augmented flamelet database. The statistics are collected over two flow-through times.}
         \label{fig_augmentedFlameletAppendix}
 \end{figure}

This section reports a comparison of the measured and computed radial profiles for temperature and select species mass fractions at $15D$, $30D$, and $45D$ downstream of the jet exit. The computed results are shown for the reference simulation using ISAT with the detailed mechanism and PPAC-ISAT using two different databases. Figure \ref{fig_ponlineAppendix} shows results for the PPAC-ISAT simulation using reduced models generated from partial online flamelet database, wherein the regions deemed problematic utilized compositions extracted from the online simulation. The results for the PPAC-ISAT simulation using the augmented flamelet database are shown in Fig. \ref{fig_augmentedFlameletAppendix}. 

Comparing the results shown for the original flamelet database in Fig. \ref{fig_flamelet} and the results for the partial online flamelet database and the augmented flamelet database shown in Figs. \ref{fig_ponlineAppendix} and \ref{fig_augmentedFlameletAppendix}, we observe improved agreement with the reference simulation using ISAT with the detailed mechanism. In addition to the improved agreement shown at $x/D=15$ in Figs. \ref{fig_ponline} and \ref{fig_augmentedFlamelet}, we see a closer match with the reference simulation at $x/D=30$.

\begin{figure}
 	\centering
 	\subfloat[Flamelet database]{%
 		\resizebox*{0.4975\textwidth}{!}{\begin{tikzpicture}
		\node[inner sep=0pt] (A) {\includegraphics{conserve_flamelet.eps}};
		\node[black,rotate=90,font=\fontsize{21}{0}\selectfont] (C) at ($(A.west)!-.035!(A.east)$) {$\mathbf{\hat{\varepsilon}_{X}}$};
		\end{tikzpicture}}
 		}
 	\subfloat[Augmented flamelet database]{%
 		\resizebox*{0.4975\textwidth}{!}{\begin{tikzpicture}
		\node[inner sep=0pt] (A) {\includegraphics{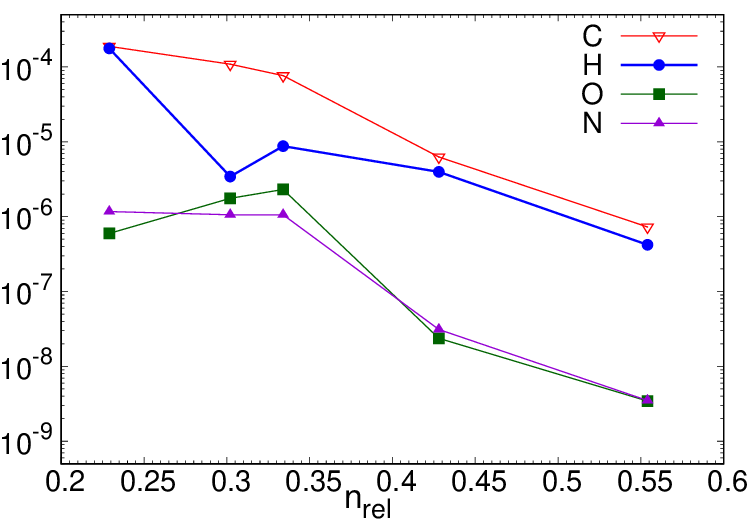}};
		\node[black,rotate=90,font=\fontsize{21}{0}\selectfont] (C) at ($(A.west)!-.035!(A.east)$) {$\mathbf{\hat{\varepsilon}_{X}}$};
		\end{tikzpicture}}
 		}
 	\caption{The elemental conservation errors for PPAC-ISAT with two different databases as a function of the relative number of species. Note that the range of the axes in the two plots is not identical.}
 	\label{fig_Augcons}
 \end{figure}

\section{Cartesian configuration results: Augmented flamelet database}
\label{sec:app_b}
For completeness, we provide analogous results to those presented in section \ref{ssec:cartesian}, for PPAC and PPAC-ISAT using reduced models generated with the augmented flamelet database. As before, we examine five different reduction thresholds of $5\times 10^{-4}$, $10^{-4}$, $5\times10^{-5}$, $10^{-5}$, and $10^{-6}$. Figure \ref{fig_Augcons} shows the elemental conservation errors as a function of the relative number of species for the augmented flamelet database cases on the right. The results for the original flamelet database are shown on the left for reference. Overall, we observe decreasing elemental conservation errors with increasing relative number of species. This can be attributed to a lower errors incurred in the conversion process going from progressively less reduced representations to the detailed representation. The conservation errors with the augmented flamelet database are also approximately bounded by $10^{-4}$, and are deemed to be acceptable. \par

\begin{figure}
 	\centering
 	\subfloat[Flamelet database]{%
 		\resizebox*{0.4975\textwidth}{!}{\includegraphics{combined_inc_T_flamelet.eps}}
 		}
 	\subfloat[Augmented flamelet database]{%
 		\resizebox*{0.4975\textwidth}{!}{\includegraphics{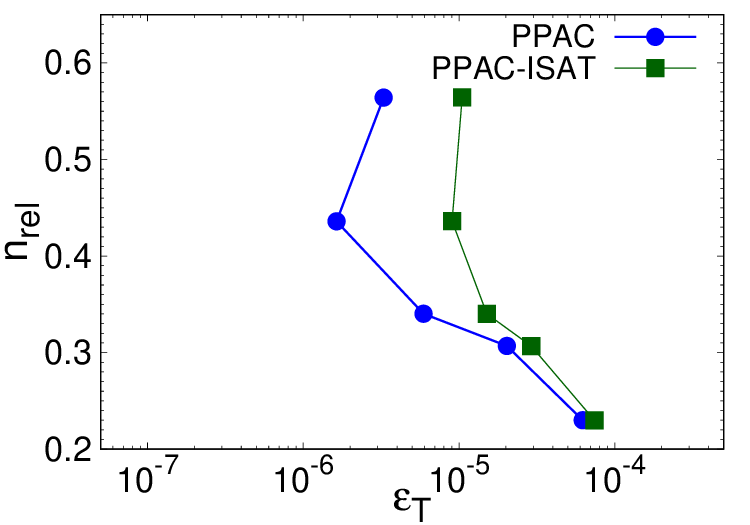}}
 		}
 	\caption{The relative number of species as a function of the incurred error in temperature for PPAC and PPAC-ISAT with two different databases. Note that the range of the axes in the two plots is not identical.}
 	\label{fig_Auginc}
 \end{figure}

 Figure \ref{fig_Auginc} shows the relative number of species as a function of the incurred error in temperature for both PPAC and PPAC-ISAT using reduced models generated with the augmented flamelet database on the right. The plot for original flamelet database is repeated on the left for ease of comparison. Using the same argument detailed in section \ref{ssec:cartesian}, we can determine the relative contribution of the reduction and tabulation errors by comparing the results for PPAC and PPAC-ISAT with the augmented flamelet database. Specifically, we can infer that the reduction error dominates for the largest two reduction thresholds. The tabulation error dominates for the remaining cases using more stringent reduction thresholds.  

\end{document}